\documentclass[12pt]{article}
\usepackage[latin1]{inputenc}
\usepackage{amsfonts}
\usepackage{amsxtra}
\usepackage[margin=3cm]{geometry}
\usepackage{color}
\usepackage{hyperref}
\hypersetup{linktocpage=true}
\usepackage{graphicx}
\usepackage{mathrsfs}
\usepackage{multirow}
\usepackage{tikz}
\usepackage{amsmath, latexsym, cite, amssymb, slashed}
\usepackage{bbm}
\usepackage{pdfpages}
\usepackage{centernot}
\usepackage{multirow}
\usepackage{varwidth}
\usepackage{slashbox}
\usepackage{amsthm}
\usepackage{xypic}

\usepackage[toc,page]{appendix}
\usepackage{bookmark}

\usepackage{slashed}
\usepackage{cancel}

\numberwithin{equation}{section}
\newcommand{\beq}{\begin{equation}}
\newcommand{\eeq}{\end{equation}}
\def\be {\begin{equation}}
\def\ee {\end{equation}}
\def\ba#1\ea{\begin{align}#1\end{align}}
\def\baed#1\eaed{\begin{aligned}#1\end{aligned}}
\def\bged#1\eged{\begin{gathered}#1\end{gathered}}
\def\bea{\begin{eqnarray}}
\def\eea{\end{eqnarray}}

\def\a{\alpha}

\def\e{\epsilon}
\def\ve{\varepsilon}

\def\F{\Phi}
\def\g{\gamma}
\def\G{\Gamma}

\def\l{\lambda}
\def\L{\Lambda}
\def\m{\mu}
\def\n{\nu}
\def\o{\omega}
\def\O{\Omega}

\renewcommand{\t}{\theta}

\def\s{\sigma}

\def\z{\zeta}

\def\cN{\mathcal{N}}

\def\Re{\text{Re}}
\def\Im{\text{Im}}
\def\Tr{\text{Tr}}

\def\ch{\text{ch}}

\let\foo\bar
\renewcommand{\bar}[1]{ {\foo{  #1} }{} }

\newlength{\dhatheight}

\newcommand{\eq}[1]{\begin{equation}\begin{split}#1\end{split}\end{equation}}

\newcommand{\al}[1]{\begin{align}#1\end{align}}

\newcommand{\arxth}[1]{\href{http://arxiv.org/abs/hep-th/#1}{[{\tt hep-th/#1}]}}
\newcommand{\arx}[1]{[\href{http://arxiv.org/abs/#1}{\tt #1}]}

\newcommand{\acal}{\mathcal{A}}

\newcommand{\dcal}{\mathcal{D}}

\newcommand{\fcal}{\mathcal{F}}

\newcommand{\hcal}{\mathcal{H}}

\newcommand{\kcal}{\mathcal{K}}
\newcommand{\lcal}{\mathcal{L}}

\newcommand{\ncal}{\mathcal{N}}

\newcommand{\pcal}{\mathcal{P}}
\newcommand{\scal}{\mathcal{S}}
\newcommand{\tcal}{\mathcal{T}}

\newcommand{\rcal}{\mathcal{R}}
\newcommand{\qcal}{\mathcal{Q}}

\newcommand{\rbb}{\mathbbm{R}}

\newcommand{\obb}{\mathbbm{1}}

\newcommand{\ea}{\bigwedge\nolimits^{\!\bullet} T^*}

\newcommand{\p}{\partial}
\newcommand{\we}{\widetilde{\eta}}
\newcommand{\vol}{\text{vol}}

\renewcommand{\Re}{\text{Re}\;}
\renewcommand{\Im}{\text{Im}~}

\renewcommand{\ch}[1]{\check{#1}}

\newcommand{\ti}[1]{\tilde{#1}}

\newcommand{\fs}{{(\star\mathcal{F})}}
\newcommand{\cg}{\check{\gamma}}

\renewcommand{\ve}{\varepsilon}

\newcommand{\cellbreak}[2][l]{%
  \begin{tabular}[#1]{@{}c@{}}#2\end{tabular}}

\numberwithin{equation}{section}

\begin{document}
%%%%%%%%%%%%%%%%%%%%%%%%%%%%%%%%%%%%%%%%%%%%%%%%

\baselineskip=16pt
\setlength{\parskip}{6pt}

\begin{titlepage}
\begin{minipage}{0.5\textwidth}
{\color{white} a}
\end{minipage}
\begin{minipage}{0.5\textwidth}
\flushright  Imperial/TP/17/HT/01
\vspace*{-.65\baselineskip}
\flushright IPhT-t17/119
\end{minipage}

%\begin{flushright}
%\parbox[t]{1.4in}{
%\flushright Imperial/TP/17/HT/01
%\vspace*{-.65\baselineskip}
%\flushright IPhT-XXXX }
%\end{flushright}

\begin{center}

\vspace*{1.7cm}

{\LARGE \bf  Supersymmetric branes and instantons on curved spaces}

\vskip 1.6cm

\renewcommand{\thefootnote}{}

\begin{center}
 \normalsize
 Ruben Minasian$^{\,a,b}  $, Dani\"el Prins$^{\,a,c,d}$, Hagen Triendl$^{\, e}$
\end{center}
\vskip 0.1cm
 {\sl $^a$ Institut de physique th\'eorique, Universit\'e Paris Saclay, CNRS, CEA \\F-91191 Gif-sur-Yvette,
France}
\vskip 0.1cm
{\sl $^b$ School of Physics, Korea Institute for Advanced Study, Seoul 130-722, Korea}
\vskip 0.1cm
{\sl $^c$ Dipartimento di Fisica, Universit\`{a} di Milano-Bicocca,\\
 I-20126 Milano, Italy}
\vskip 0.1cm
{\sl $^d$ INFN, sezione di Milano-Bicocca, I-20126 Milano, Italy}
\vskip 0.1cm
{\sl $^d$ Department of Physics, Imperial College London, London SW7 2AZ, UK}
\vskip 0.2 cm
{\textsf{ruben.minasian@cea.fr}} \\
{\textsf{daniel.prins@cea.fr}}\\
{\textsf{h.triendl@imperial.ac.uk}}
\end{center}

\vskip 1.5cm
%\addtocounter{footnote}{-1}
\renewcommand{\thefootnote}{\arabic{footnote}}

\begin{center} {\bf ABSTRACT } \end{center}
We discuss non-linear instantons in supersymmetric field theories on curved spaces arising from D-branes. Focussing on  D3-branes and four-dimensional field theories, we derive the supersymmetry conditions and show the intimate relation between the instanton solutions and the non-linearly realized supersymmetries of the field theory. We demonstrate that field theories with non-linearly realized supersymmetries are coupled to supergravity backgrounds in a similar fashion as those with linearly realized supersymmetries, and provide details on how to derive such couplings from a type II perspective.

\end{titlepage}

\newpage
%\noindent\rule{\textwidth}{.1pt}		
\tableofcontents
\vspace{20pt}
%\noindent\rule{\textwidth}{.1pt}

\renewcommand\arraystretch{1.2}
\setcounter{page}{1}
\setlength{\parskip}{9pt}
\section{Introduction}

In recent years new localization techniques have made it possible to gain deeper insights into supersymmetric field theories by computing their partition function on curved spaces (see \cite{Pestun:2016zxk} for a review on the subject). There are two questions of major interest for the  localization programme:
\begin{enumerate}
\item On which curved spaces can a supersymmetric field theory be realized in a supersymmetric fashion?
\item What are the classical supersymmetric solutions of a supersymmetric field theory on that curved space?
\end{enumerate}
The first problem has been extensively studied by coupling supersymmetric field theories to background off-shell supergravities and by subsequently tuning the background fields of the supergravity multiplet in order to  compensate the appearance of curvature terms in the supersymmetry conditions.
A fairly thorough study of admissible four-dimensional spaces preserving four or less supercharges is available: see \cite{fs,ktz,dfs, df, cmktz, liu, st, js, cfdz} for early works on this subject. The second problem naturally leads to the study of instanton solutions of supersymmetric theories on a given space, which will be the focus of this paper.

Supersymmetric field theories on curved spaces naturally arise in string and M-theory, usually realized by systems of calibrated branes in a given flux background. Embeddings of supersymmetric field theories in string theory are interesting for various reasons. Brane systems offer a UV-completion for supersymmetric field theories, and the geometric properties and string dualities of these brane systems often reveal non-perturbative properties of the field theory. This work aims at a deeper understanding of supersymmetric field theories realized as brane systems in flux compactifications of string theory.

The relationship between the two setups, i.e. the branes in flux compactifications of string and M-theory and the  description of a supersymmetric field theory coupled to off-shell supergravity, has been recently clarified in \cite{triendl} and further studied in \cite{Maxfield:2015evr} for the case of M5-branes.
In the limit where physics is determined to leading order by the worldvolume action of the brane, the bulk fields of the string or M-theory become non-dynamical.
These are then naturally identified with the background fields in the supergravity multiplet coupled to the supersymmetric field theory living on the worldvolume. In this way, a single stack of D3- or M5-branes naturally couples to conformal $\ncal=4$ supergravity \cite{ht}.

One major feature of brane systems in string and M-theory is that they admit additional supersymmetries that are non-linearly realized on the brane worldvolume \cite{Hughes:1986dn}.
The fact that these supersymmetry transformations are realized non-linearly on the worldvolume is reminiscent of spontaneous partial supersymmetry breaking \cite{Antoniadis:1995vb,Bagger:1996wp} and explains the non-linear nature of the DBI action \cite{Rocek:1997hi}. It also naturally leads to non-commutative systems within string theory \cite{sw}. These non-linearly realized supersymmetries are crucial for the understanding of supersymmetric brane systems in flux compactifications.

Since the supersymmetry conditions on branes depend on the worldvolume gauge field via its field strength $\cal F$, this worldvolume flux is constrained by supersymmetry. As always, the situation is better understood in  compactifications without bulk fluxes. The study of supersymmetric branes with worldvolume flux inside special holonomy manifolds shows that the supersymmetry conditions usually decompose into a calibration condition on the submanifolds wrpapped by branes, which is independent of the worldvolume fields, and a, generally non-linear,  instanton equation for the worldvolume flux \cite{mmms}. In more general flux backgrounds such splitting of supersymmetry conditions into separate constraints on geometry of wrapped branes and on gauge-theoretic conditions  seems a priori unlikely. It is the aim of this paper to understand the interplay of background bulk fluxes and worldvolume flux $\cal F$ in the field theory.

As we will see, in the presence of general worldvolume flux, supersymmetric solutions on the worldvolume naturally involve the non-linearly realized supersymmetries and lead to non-linear instantons. In order to understand these solutions on curved spaces, we have to analyze the coupling of supergravity backgrounds to the full non-linear theory, including the non-linearly realized supersymmetries. We will study this problem for a Euclidean D3-brane in a general (not necessarily compact) flux background and show that we can describe that system by coupling the corresponding $\cN=4$ worldvolume field theory to $\cN=8$ background supergravity.\footnote{The case of Lorentzian D3-branes can be discussed analogously. In that case the $SU^*(8)$ R-symmetry group is replaced by its compact counterpart $SU(8)$.} Employing this language, we then show that the supersymmetry conditions still split into a supersymmetry condition for the $\cN=8$ background supergravity fields and a non-linear instanton equation for the worldvolume flux. If solutions to both equations preserve the same supersymmetry, the resulting system is overall supersymmetric. Solutions with less supersymmetry can also be analyzed in similar fashion using intersecting brane systems.

This work is organized as follows. In Section \ref{sec:basics} we review the supersymmetry conditions of D-branes in type II backgrounds, with an emphasis on the linear and non-linear supersymmetries of the worldvolume theory on the brane. In Section \ref{sec:instanton} we will discuss linear and non-linear instanton solutions, and clarify the relation between non-linear instantons and non-linear supersymmetry. In Section \ref{2d}, we discuss the coupling of field theories with non-linear supersymmetries to supergravity backgrounds. Conventions, technical details
and the relation to Licherowitcz theorem is relegated to the appendices.

\section{D-brane field theories} \label{sec:basics}

In this section, we shall briefly review the basics of the bulk and brane supersymmetries (linear and non-linear) and set up some of our notations. More details on the latter can be found in Appendix \ref{sec:C&I}.

\subsection{Supersymmetric D-branes in type IIB flux backgrounds}
Since our main focus will be on four-dimensional theories arising form D3 branes, we shall concentrate on  supersymmetric type IIB flux backgrounds.
A (bosonic) type IIB supersymmetric flux background is a profile for the IIB fields $(g, \phi, F, H)$ such that the supersymmetry variation of the gravitino doublet $\Psi_M= (\Psi_M^1, \Psi_M^2)$
\begin{equation}\label{eq:gravitinovar}
\delta_{\hat \varepsilon} \Psi_M =  \hat{D}_M \hat{\ve}
= \nabla_M \hat \varepsilon + \tfrac18 H_{MNP} \Gamma^{NP} {\cal P} \hat \varepsilon + \tfrac{1}{16} e^{\phi} \sum_n \tfrac{1}{(2n-1)!} F_{M_1 \dots M_{2n-1}} \Gamma^{M_1 \dots M_{2n-1}} \Gamma_M {\cal P}_{n} {\hat \varepsilon} \ ,
\end{equation}
and the variation of the dilatino $\chi= (\chi^1, \chi^2)$
\begin{equation} \label{eq:dilatinovar}
 \delta_{\hat \varepsilon} \chi = \hat{D} \hat{\varepsilon} = ( (\partial_M \phi)\Gamma^M {\hat \varepsilon} +\tfrac{1}{12} H_{MNP} \Gamma^{MNP}{\cal P}) {\hat \varepsilon} - \tfrac18 e^{\phi} \sum_n  \tfrac{6-2n}{(2n-1)!}  F_{M_1 \dots M_{2n-1}} \Gamma^{M_1 \dots M_{2n-1}} {\cal P}_{n} {\hat \varepsilon} \
\end{equation}
vanish for some spinor doublet $\hat \varepsilon= (\hat \varepsilon^1, \hat \varepsilon^2)$, where both ${\hat \ve^1}$ and ${\hat \ve^2}$ have positive chirality. Here
\begin{equation}
{\cal P}   = \left(\begin{array}{cc} 1 & 0 \\ 0 & -1 \end{array}\right) \ , \qquad
{\cal P}_n = \left(\begin{array}{cc} 0 & 1 \\  (-1)^{n} & 0 \end{array}\right) \ ,
\end{equation}
generate the symmetry group ${\rm SL}(2, \mathbb{R})$ of type IIB supergravity.
Note that $\cal P$, ${\cal P}_n$ and ${\cal P}_{n+1}$ anti-commute with each other and obey the commutation relations
\begin{equation} \label{eq:Pn_comm}
 [{\cal P}, {\cal P}_n] = 2 {\cal P}_{n+1} \ , \qquad [{\cal P}_n, {\cal P}_{n+1}] = 2 (-1)^{n+1} {\cal P} \ .
\end{equation}
Generically we will not require a background to obey the equations of motions. In the following we will abbreviate the vanishing of \eqref{eq:gravitinovar} and \eqref{eq:dilatinovar} by
\begin{equation} \label{eq:susycond1}
 \hat{D}_M {\hat \varepsilon}= 0 \ , \qquad \hat D {\hat \varepsilon}= 0 \ ,
\end{equation}
respectively.

The supersymmery condition of a D$p$-brane (also referred to as kappa-symmetry) is described by the equation \cite{kappasym}
\begin{equation} \label{eq:kappa_symm}
 \Gamma {\hat \varepsilon}  = {\hat \varepsilon} \ ,
\end{equation}
and the amount of the unbroken supersymmetry is determined by the dimension of the space of its solutions. $\Gamma$ is a traceless Hermitian matrix which squares to one,
and for a Euclidean D$p$-brane in type IIB string theory takes the form:\footnote{We will only consider the Euclidean scenario in this paper, where the D$p$-brane does not fill out the time direction. For Lorentzian D$p$-branes, $\G$ comes with an additional factor of $-i$ and there is a minus sign in front of the determinant.}
\begin{equation} \label{eq:kappa_projector}
 \Gamma(\fcal) = \frac{i \, \sqrt{{\rm det}(g)} }{\sqrt{{\rm det}(g+ {\cal F})}} \sum_{2l+s = p+1} \tfrac{1}{l! s! 2^l} \epsilon^{n_1\dots n_{2l} m_{1} \dots m_s } {\cal F}_{n_1 n_2} \dots {\cal F}_{n_{2l-1} n_{2l}} \gamma_{m_1 \dots m_s} {\cal P}_{\frac12 s + 1} \ ,
\end{equation}
where $m, n, ... \in \{1,...,p+1\}$ are flat worldvolume indices, $M, N, ... \in \{ 0, ..., 9\}$ curved spacetime indices, and $\gamma_m$ are the pullbacks of the spacetime gamma-matrices $\G_M$: $\gamma_m = E_m^M  \Gamma_M$. In order to differentiate between the two, we will occasionally refer to \eqref{eq:gravitinovar}, \eqref{eq:dilatinovar} as the closed string supersymmetry conditions and to \eqref{eq:kappa_symm} as the open string supersymmetry conditions.

Let $\scal$ be the submanifold of the ten-dimensional spacetime wrapped by the D$p$-brane.
On ${\cal S}$ we can work locally and split the tangent bundle of the ten-dimensional spacetime into tangent and normal  directions to ${\cal S}$; normal directions will be labeled by indices $a, b,...\in \{0, 5, 6, .., 9\}$. We shall slightly depart from from the standard analysis of \eqref{eq:kappa_symm}, and rather than working with the ten-dimensional spinors ${\hat \varepsilon}$,  phrase our discussion in terms of  $\varepsilon$  defined as the restriction of $\hat \varepsilon$ to ${\cal S}$
\begin{equation}
\varepsilon = \hat \varepsilon \Big|_{\cal S} \ ,
\end{equation}
which can then be decomposed as a  product of ``external" spinors  $\eta$ which live on $\scal$ and ``internal" spinors $\xi$ which live on the space transverse to the brane. Of course, $\cal S$ need not be a spin manifold, and hence both $\eta$ and $\xi$ are defined only locally. We shall return to this point later. For our purposes, the local analysis is sufficient and  from now on we shall restrict \eqref{eq:kappa_symm} to $ \Gamma { \varepsilon}  = {\varepsilon}$. We similarly restrict the differential operator $\hat D_M$ to the D-brane worldvolume ${\cal S}$ so that $D_M= \hat D_M |_{\cal S}$. Trivially, we do the same for the algebraic operator $\hat D$, so that the closed string supersymmetry conditions on the worldvolume read
\begin{equation}
\label{eq:susycond2}
D_m \varepsilon= 0 \ , \qquad D  \varepsilon = 0 \ .
\end{equation}
We will refer to the first equation as the external gravitino equation.
In total, the supersymmetry conditions of type IIB also include the internal gravitino variation, which is perpendicular to the D$p$-brane and given by
\begin{equation}
\label{eq:susycond3}
D_a \ve = 0\;.
\end{equation}
We shall mostly concentrate here on \eqref{eq:susycond2}, describing the external gravitino and dilatino variations,  since these are the equations pertinent for the description of the supersymmetric gauge theory on curved spaces. The internal gravitino equation \eqref{eq:susycond3} describes the embedding of the brane into the full type IIB background.

%While the external gravitino and the dilatino equation contain information regarding the brane, the internal gravitino equation describes how to embed the brane into the IIB background. Our primary concern here is the former, so we shall not discuss the internal gravitino equation in any detail.

\subsection{Linear and non-linear supersymmetries}

For the case of vanishing worldvolume flux ${\cal F}$, the matrix $\Gamma$ given in \eqref{eq:kappa_projector} simplifies drastically. We find
\begin{equation}
\Gamma_0 \equiv \Gamma({\cal F}=0) = i \gamma_{(p+1)}{\cal P}_{(p-1)/2} \ ,
\end{equation}
and we can split the supersymmetry generators $\varepsilon$ into\footnote{Note that here we use subscripts for the transformation behavior under $\Gamma_0$, while $(p+1)$-dimensional chirality will be denoted by superscript. The difference between the two notions lies in the appearance of the ${\rm sl}(2,\mathbb{R})$ matrix ${\cal P}_{(p+1)/2}$ in $\Gamma_0$.}
\begin{equation}
 \varepsilon_\pm = \tfrac12 (1 \pm \Gamma_0) \varepsilon \ .
\end{equation}
We see that for ${\cal F}=0$ the supersymmetries $\varepsilon_+$ are preserved by the brane, while the supersymmetries $\varepsilon_-$ are broken.

On the worldvolume we can understand the calibration condition \eqref{eq:kappa_symm} as the supersymmetry variation of the gaugino $\lambda$.
After appropriate gauge choice for kappa-symmetry the physical gaugino $\lambda$ obeys \cite{kappasym}
\eq{
\Gamma_0 \lambda &= - \lambda \\
\delta \lambda &= \tfrac14 (1-\Gamma_0)(\Gamma({\cal F})-1) \varepsilon + \dots \;,
}
where the dots suppress terms depending on the worldvolume scalars.
We can express this in terms of $ \varepsilon_\pm $ as
\begin{equation}
\delta \lambda =  \delta_+ \lambda + \delta_- \lambda  \ ,
\end{equation}
where
\eq{
 \delta_+ \lambda &= \tfrac12 [\Gamma_0,\Gamma({\cal F})] \varepsilon_+ + \dots \\
 \delta_- \lambda &= - \tfrac12 \{ \Gamma_0,\Gamma({\cal F})-1 \} \varepsilon_- + \dots \ .
}

For a stack of D3-branes, the supersymmetry variation with respect to $\varepsilon_+$ reads
\begin{equation} \label{eq:linvar}
 \delta_+ \lambda = \cancel{{\cal F}} {\cal P} \ve_+ + \dots \ ,
\end{equation}
which corresponds to the standard linear supersymmetry variation of the $\ncal=4$ vector multiplet.
In contrast, the gaugino transforms under the supersymmetry variation with respect to $\varepsilon_-$, as
\begin{equation} \label{eq:nonlinvar}
 \delta_- \lambda = \left(2 - \frac14 (\ast {\cal F})_{mn} {\cal F}^{mn} \gamma_{(4)} + \sqrt{ \det( \delta + g^{-1}\fcal) } \right)  \varepsilon_- + \dots \ .
\end{equation}
For vanishing ${\cal F}$ this variation clearly is non-zero, indicating that these supersymmetries are spontaneously broken in the vacuum on the worldvolume.
The variation under $\varepsilon_-$ corresponds to the non-linear supersymmetry transformations on the D3-brane worldvolume \cite{Bagger:1996wp,Rocek:1997hi}. These supersymmetry variations are particular to D-branes actions and the DBI action.
Although these supercharges are always broken for the vacuum with $\fcal =0$, for general configurations of ${\cal F}$, they allow for more general, non-linear instanton solutions on the worldvolume. Thus these non-linearly realized supersymmetries are crucial for understanding the spectrum of supersymmetric states on the worldvolume.

Note that the supersymmetry variation with respect to $\varepsilon_+$ is only linear for D$p$-branes with $p\le 4$. For $p\ge 5$ an additional cubic term appears in \eqref{eq:linvar}, which means the Yang-Mills supersymmetry variation receives an $\alpha'$ correction. We will discuss instantons for the case of D5-branes below in Section \ref{D5inst}.

\section{Instantons on D3-branes}\label{sec:instanton}
Let us now turn our attention to instanton solutions for the worldvolume theory of an Euclidean D-brane. Supersymmetric instantons are characterized by $\delta \lambda=0$ for some supersymmetry parameter $\varepsilon$ and some non-zero profile for the worldvolume flux ${\cal F}$.
We will discuss four-dimensional field theories related to a D3-brane in a supersymmetric flux background.
We will first consider the case where the supersymmetries preserved by the instanton are those linearly realized on the worldvolume, i.e.\ $\varepsilon_-=0$, and show that this is exactly the case of linear instantons, namely when the condition for preserving supersymmetry is given by the ordinary  Hermitian Yang-Mills equation. Subsequently we will discuss the instanton equation for general $\varepsilon$. We will close the section by comments on the instanton equation for D5-branes.

At various points in this section, we will decompose ten-dimensional Killing spinors $\hat \ve$ into non-vanishing components parallel and perpendicular to the brane. The rest of the discussion will be phrased in terms of $\ve$ which can be written as a  product of external and internal spinors. Our discussion is purely local, and we shall not worry about $\ve$ being globally well-defined. We tacitly assume that the manifold $\scal$, wrapped by the brane,  allows for a local $SU(2)$-structure. Cases when $\scal$ is not an $SU(2)$-structure manifold are studied in the literature and are well-understood. Notable examples are given by the D3-brane wrapping a generic K\"{a}hler manifold or $S^4$. There are two ways in which our decomposition can go wrong. Firstly, one need not have a spin-structure on $\scal$, in which case one should decompose $\ve$ into ``charged" $Spin^c(4)$ spinors on $\scal$. This is the case for $U(2)$-holonomy (K\"ahler) manifolds. Secondly, the spinors along or perpendicular to the brane may have local zeros, as is the case for D3-branes on $S^4$. We shall refer to e.g.  \cite{ktz, dfs} for in-depth discussion of these issues, and from now one carry on with a local analysis in terms of $\ve$.

For now we will ignore the coupling of the worldvolume supersymmetries to the bulk fields and the possible supersymmetry breaking induced by these couplings. We will come back to this point in section \ref{2d}.

\subsection{Hermitian Yang-Mills}\label{hym}
Let us start by studying instanton solutions whose supersymmetries are linearly realized. In other words, let us assume for now that $\ve_-=0$,
such that $\G_0 \ve = \ve$, with in our case of D3-branes,
\eq{
\G_0 = i \g_{(4)} \pcal_1 \;.
}
Then we see already from \eqref{eq:linvar} that supersymmetry implies the Hermitian Yang-Mills equation\footnote{This can be derived straightforwardly from imposing $\G \ve = \G_0 \ve = \ve$.}
\begin{equation} \label{eq:lininst}
\frac12 \fcal_{mn} \g^{mn} \varepsilon = 0 \ .
\end{equation}
The supersymmetry condition \eqref{eq:kappa_symm} implies that we can rewrite the Killing spinor $\varepsilon$ as
\eq{
\varepsilon_+ = (\ve^1_+, -i \g_{(4)} \ve^1_+) \;.
}
In order to solve the Hermitian Yang-Mills equation, we decompose the ten-dimensional spinor $\ve^1_+$ via $Spin(1,9) \rightarrow Spin(1,5) \times Spin(4)$ as
\eq{
\ve^1_+ = \xi^+_\a \otimes \eta^+_\a + \xi^-_\a \otimes \eta^-_\a \;,
}
with $\xi^\pm_\a$ chiral spinors of $Spin(1,5)$ perpendicular to the brane, $\eta^\pm_\a$ chiral spinors of $Spin(4)$ on the brane. Generically, $\a \in \{1,..,4\}$ such that we have 16 supercharges.\footnote{Note that unlike for $Spin(1,3)$, charge conjugation does not change chirality, hence $\eta^+$ is independent from $\eta^-$. Thus the (Majorana-Weyl) reality condition of $\ve$ leads to some constraints on the decomposed spinors, i.e., we could also have written
\eq{
\ve^1_+ = \xi^+_\a \otimes \eta^+_\a + \xi^-_\a \otimes \eta^-_\a + \text{c.c.} \;, \qquad \a \in \{1,2\} \;.
}
}
Since $\xi^\pm_\a$ are independent for each $\a$, each $\eta_\a^\pm$ will yield a copy of the instanton equation, but with respect to a possibly different $SU(2)$-structure determined by $\eta^\pm_\a$. We will solve for a specific $\a$ and note that the instanton solution is the intersection of the solutions to each seperate equation. Thus, \eqref{eq:lininst} reads
\eq{\label{eq:inst1}
\xi^+ \otimes  \cancel{\fcal}  \eta^+ + \xi^- \otimes  \cancel{\fcal}  \eta^- &= 0 \; .
}
which only admits a non-trivial solution for $\xi^-=0$ or $\xi^+=0$. We will consider $\xi^- =0$, such that
\begin{equation} \label{eq:4dchiral}
\gamma_{(4)} \varepsilon = \varepsilon \ ,
\end{equation}
The normalized chiral spinor $\eta \sim \eta^+$ defines on the worldvolume an $SU(2)$-structure via
\eq{\label{su2}
- i \we^c \g_{mn} \eta &= J_{mn} \\
\we \g_{mn} \eta &= \O_{mn} \;.
}
Thus \eqref{eq:lininst} reduces to
\eq{
\fcal_{mn} J^{mn} = \fcal_{mn}\O^{mn} &= 0 \;,
}
which is equivalent to stating that $\fcal = \star_4 \fcal$ with respect to the metric defined by the $SU(2)$-structure.
One consequence of \eqref{eq:4dchiral} is that the DBI-like term simplifies to
\begin{equation} \label{eq:detsimplify}
 \sqrt{ \det( \delta + g^{-1}\fcal) }  = 1+ \frac14 \fcal_{mn} \fcal^{mn} \ .
\end{equation}
Had we instead considered $\xi^+ =0$ we would have found an anti-selfdual worldvolume flux.

Conversely, we can show that any spinor $\varepsilon$ solving both \eqref{eq:kappa_symm} and \eqref{eq:lininst} has to obey $\G_0 \varepsilon = \varepsilon$, or in other words $\varepsilon_-=0$.
For this, note
\eq{ \label{eq:gammaFdecomp}
(\G- 1) \ve = (\G_0 - 1) \ve - \cancel{\fcal} \G_0 \pcal \ve + \left( \sqrt{ \det( \delta + g^{-1}\fcal) } - 1-  \frac14 \fcal_{mn} \fcal^{mn}\g_{(4)} \right) \ve \ .
}
The second term is obviously zero due to selfduality \eqref{eq:lininst}, while the third term vanishes due to \eqref{eq:detsimplify}, hence we find
\eq{
\G_0 \ve= \ve\ .}
In similar fashion we can use \eqref{eq:gammaFdecomp} to show that the linear instanton equation \eqref{eq:lininst} and $\G_0 \varepsilon = \varepsilon$ together imply \eqref{eq:kappa_symm}.

In total we have shown that any two of the three following statements imply the third:\footnote{Note that this does not contradict the conclusions of \cite{mmms}, where one finds non-linear instanton equations on calibrated sumbanifolds of manifolds of special holonomy. The general solutions of  $\Gamma(\fcal) \varepsilon = \varepsilon$ reduce to solutions of $\Gamma_0\varepsilon = \varepsilon$ not only by dropping $\cal F$ but also setting to zero a phase, which is responsible for the deformation of the HYM equations.}
\begin{enumerate}
\item $\varepsilon$ fulfills the instanton equation $\Gamma(\fcal) \varepsilon = \varepsilon$,
\item $\varepsilon$ corresponds to a linearly realized supersymmetry, i.e.\ $\Gamma_0\varepsilon = \varepsilon$,
\item $\varepsilon$ satisfies the Hermitian Yang-Mills equation $\cancel{\fcal} \varepsilon = 0$.
\end{enumerate}
Thus linear instantons are linked to the corresponding supersymmetry being linearly realized on the worldvolume. In the following we will discuss non-linear instantons that are related to general supersymmetries on the worldvolume of the brane.

\subsection{Non-linear instantons}\label{nli}
We now want to turn to the more general case where both $\varepsilon_+$ and $\varepsilon_-$ are non-zero for the supersymmetric instanton configuration on the Euclidean D3-branes, and study the case
\begin{equation}
\delta \lambda = \delta_+ \lambda + \delta_- \lambda = 0 \ ,
\end{equation}
which corresponds to the general solution to \eqref{eq:kappa_symm}. As this equation is highly non-linear in the worldvolume flux $\cal F$, we call these configurations non-linear instantons, following \cite{mmms}.

For any D$p$-brane we can compose the spinors $\varepsilon_+$ and $\varepsilon_-$ into a doublet. In this way, we find from \eqref{eq:kappa_symm} that
\begin{equation}
\tfrac12
\left( \begin{aligned}
\{ \Gamma , \Gamma_0 \} && -[\Gamma , \Gamma_0 ] \\ [\Gamma , \Gamma_0 ] && -\{ \Gamma , \Gamma_0 \}
\end{aligned} \right)
\left( \begin{array}{c}
\varepsilon_+ \\ \varepsilon_-
\end{array} \right)
=
\left( \begin{array}{c}
\varepsilon_+ \\ \varepsilon_-
\end{array} \right) \ .
\end{equation}
Using \eqref{eq:kappa_projector}, this reduces to
\begin{equation}\label{eq:nonlininstgen}
\sqrt{\frac{{\rm det}(g)}{{\rm det}(g+ {\cal F})}}
\left( \begin{array}{cc}
\cancel{\cosh({\cal F})}  & \cancel{\sinh({\cal F})} {\cal P}\\
-\cancel{\sinh({\cal F})} {\cal P} & -\cancel{\cosh({\cal F})}
\end{array} \right)
\left( \begin{array}{c}
\varepsilon_+ \\ \varepsilon_-
\end{array} \right)
=
\left( \begin{array}{c}
\varepsilon_+ \\ \varepsilon_-
\end{array} \right) \ ,
\end{equation}
where in the power series of $\cosh$ and $\sinh$ the argument ${\cal F}$ is multiplied by wedge products. For a D3-brane we have
\begin{equation}
\sinh({\cal F}) = {\cal F} \ , \qquad \cosh({\cal F}) = 1 + \tfrac12 {\cal F}\wedge {\cal F} \ ,
\end{equation}
so that \eqref{eq:nonlininstgen} simplifies to
\eq{ \label{eq:nonlininst}
\sqrt{\frac{{\rm det}(g)}{{\rm det}(g+ {\cal F})}}
\left( \begin{array}{cc}
1 + \tfrac18 \epsilon^{mnpq}{\cal F}_{mn}{\cal F}_{pq} \gamma_{(4)}  & \cancel{\cal F} {\cal P}\\
 -\cancel{\cal F} {\cal P}& -1 - \tfrac18 \epsilon^{mnpq}{\cal F}_{mn}{\cal F}_{pq} \gamma_{(4)}
\end{array} \right)
\left( \begin{array}{c}
\varepsilon_+ \\ \varepsilon_-
\end{array} \right)
=
\left( \begin{array}{c}
\varepsilon_+ \\ \varepsilon_-
\end{array} \right) \ .
}
This is the general non-linear instanton equation.

Note that \eqref{eq:nonlininstgen} only involves gamma matrices on the brane worldvolume, and is therefore diagonal on the normal bundle. As a consequence, the internal spinors in the decomposition of $\ve_\pm$ should be equivalent. We thus consider
\eq{\label{spindec10}
(\ve_+)^1 &=  \xi^+_\a \otimes \eta^+_\a + \xi^-_\a \otimes \eta^-_\a \\
(\ve_-)^1 &=  \xi^+_\a \otimes  \z^+_\a  + \xi^-_\a \otimes  \z^-_\a  \;.
}
Unlike for the HYM case, solutions do exist with both $\xi^+ \otimes \eta^+$ and $\xi^- \otimes \eta^-$ non-trivial. As when taking the sum over $\a$ into account, the inclusion of both $\eta^\pm$ leads to two sets of equations. The two sets are similar (up to sign changes due to $\g_{(4)}$), but may possibly involve different $SU(2)$-structures,  which are determined by the relation between $\eta^+$ and $\eta^-$. Such solutions are a subset of the ones analyzed here, and are relevant when coupling field theories to curved backgrounds as determined by the Killing spinor equations (as we will discuss in the next section) since both $\eta^+$ and $\eta^-$ make an appearance in the gravitino variations. Having said this, we will demonstrate the solution in the case where we consider just a single term with external spinors $\eta^+$ and $\z^+$.

Given a normalized positive-chirality spinor $\eta$ of $Spin(4)$, any spinors $\psi_\pm$ of $Spin(4)$ may locally be written as
\eq{\label{spindec}
\psi^+ &= a \eta + b \eta^c \\
\psi^- &= c_m \g^m \eta \;.
}
Let us for the moment restrict our attention to the case where we consider
\eq{\label{spindec2}
\eta^+ &= a \eta + a^*  \eta^c \\
\z^+   &= b \eta + b^* \eta^c \;.
}
Inserting \eqref{spindec10}, \eqref{spindec2} into \eqref{eq:nonlininst} and multiplying both sides by  both $\vol_4 \tilde{\eta}^c$ and $\vol_4 \tilde{\eta}$, we  find the following set of equations:
\eq{\label{mmms1}
  a   i \fcal \wedge J - a^* \fcal \wedge \O^* &= b   \left( (\sqrt{ \det( \delta + g^{-1}\fcal) } + 1) \vol_4 + \frac12 \fcal \wedge \fcal \right) \\
- a^* i \fcal \wedge J + a   \fcal \wedge \O   &= b^* \left( (\sqrt{ \det( \delta + g^{-1}\fcal) } + 1) \vol_4 + \frac12 \fcal \wedge \fcal \right) \\
- b   i \fcal \wedge J + b^* \fcal \wedge \O^* &= a   \left( (\sqrt{ \det( \delta + g^{-1}\fcal) } - 1) \vol_4 - \frac12 \fcal \wedge \fcal \right) \\
  b^* i \fcal \wedge J - b   \fcal \wedge \O   &= a^* \left( (\sqrt{ \det( \delta + g^{-1}\fcal) } - 1) \vol_4 - \frac12 \fcal \wedge \fcal \right) \;,
}
which can be simplified to find
\eq{\label{mmmsresult}
\fcal \wedge \O &= 0  \\
\fcal \wedge J &=  k \left(\vol_4 + \frac12 \fcal \wedge \fcal \right) \;.
}
We have defined here
\eq{
k = \frac{ 2 (\frac{b}{a i})}{1 -(\frac{b}{a i})^2} \;.
}
The conditions \eqref{mmmsresult} are exactly the four-dimensional non-linear instanton equations of  \cite{mmms}.\footnote{ Our variables may be related to those of \cite{mmms} via
\eq{
a = i e^{- \frac12 i \a} \cos \frac{\a}{2} \, , \quad
b = e^{- \frac12 i \a} \sin \frac{\a}{2} \, , \quad
e^{i \a} = i e^{i \t}
}
such that $k = \tan \a$. The sign difference for $\fcal^2$ is due to a different choice of chirality.}
%In order to obtain their (3.9) from (3.8), they have set $U= e^\frac{i \a} \obb$, which amounts to setting $w_+ = - w_- = 1$.}
%\Hnote{Isn't it a divisor in a CY in \cite{mmms}, i.e.\ U(2) holonomy?}

As it stands, due to the ansatz \eqref{spindec2}, this solution is not the most general. By writing \eqref{spindec2}, we have essentially ``gauge fixed" a choice of $SU(2)$-structure. One can obtain another $SU(2)$-structure, either at the level of the spinors by considering
\eq{
\left(\begin{array}{c} \eta \\ \eta^c \end{array}\right) \rightarrow X \left(\begin{array}{c} \eta \\ \eta^c \end{array}\right) ~, \quad X \in SU(2) \;,
}
or, equivalently, at the level of the forms by considering
\eq{
\left(\begin{array}{c} J \\ \Re \O \\ \Im \O \end{array}\right) \rightarrow R  \left(\begin{array}{c} J \\ \Re \O \\ \Im \O \end{array}\right) \;, \quad R \in SO(3) \;.
}
Note that such a rotation does not affect the metric; the metric is invariant under $Spin(4)\simeq SU(2) \times SU(2)$ frame transformations, of which the above $SU(2)$ is a subgroup.
%In fact, this $SU(2)$-rotation is exactly the `missing' piece of the matrix $U \in SU(2) \times U(1)$ given in \cite{mmms} (3.8).
See \cite{mmms} for more details, or compare with  \cite{aspinwall, pta} for occurrences of this phenomenon for K3 surfaces in the context of flux compactifications.

\subsection{Instantons on D5-branes}
\label{D5inst}
In this section, we will take a slight detour and consider a D5-brane rather than a D3-brane. Although the derivation of the coupling of six-dimensional instantons to supergravity backgrounds from string theory is beyond the scope of this paper, we refer to \cite{tsimpis} for analysis of such six-dimensional supergravity backgrounds. We repeat the derivation of both the ``linear" and the non-linear instanton equation for a Euclidean D5-brane. Similarly to the D3-brane case, we shall refer to taking $\ve_- = 0$ as the linear instanton equation. However, we will find that for the D5-brane, the resulting ``linear" instanton equation is actually non-linear in $\cal F$.
\\
\\
Using similar notation as before, the kappa-symmetry operator $\G$ for a D5-brane is given by
\eq{
\G &= \frac{1}{ \sqrt{ \det( \delta + g^{-1}\fcal) }} \left( \G_0 + \G_1 + \G_2 + \G_3 \right) \\
\G_0 &\equiv i \, \g_{(6)}  \, \pcal_3 \\
\G_1 &\equiv - \cancel{\fcal} \, \pcal \, \G_0 \\
\G_2 &\equiv \qcal \, \g_{(6)}  \, \G_0 \\
\G_3 &\equiv \rcal \, \g_{(6)} \, \pcal \, \G_0 \;,
}
with
\eq{\label{pqr}
%p &= \sqrt{ \det( \delta + g^{-1}\fcal) } - 1 \\
\qcal &= - \frac18 \fs_{mnpq} \fcal^{mn} \g^{pq} \\
\rcal &= - \frac{1}{3!} \frac{1}{2^3} \e^{mnpqrs} \fcal_{mn} \fcal_{pq} \fcal_{rs} \;.
}
Inserting these into the supersymmetry constraint $\G \ve = \ve $
leads to
\eq{
& \left( (  \cancel{\fcal} - \rcal \g_{(6)} \pcal + ( \sqrt{ \det( \delta + g^{-1}\fcal) } - \qcal \g_{(6)} - 1) \right) \ve_+ \\
=
& \left( (  \cancel{\fcal} - \rcal \g_{(6)}) \pcal + ( \sqrt{ \det( \delta + g^{-1}\fcal) } + \qcal \g_{(6)} + 1) \right) \ve_-\;.
}
Using $\ve_\pm = (\ve^1_\pm, \pm i \g_{(6)} \ve^1_\pm)$ then yields the non-linear instanton equations
\eq{\label{d=6nlinstanton}
(\cancel{\fcal} - \rcal \g_{(6)}) \ve_+^1 &= - (1 +  \sqrt{ \det( \delta + g^{-1}\fcal) } + \qcal \g_{(6)}) \ve_-^1 \\
 ( \sqrt{ \det( \delta + g^{-1}\fcal) } - \qcal \g_{(6)} - 1) \ve_+^1 &=  (  \cancel{\fcal} - \rcal \g_{(6)}) \ve_-^1 \;.
}
Let us now solve these non-linear instanton equations, proceeding analogously to the D3-brane case. First, we decompose the Killing spinors as
\eq{
\ve_+^1 = \xi^+ \otimes \eta^+ + \xi^- \otimes \eta^-\\
\ve_-^1 = \xi^+ \otimes \z^+   + \xi^- \otimes \z^-
}
with $\xi^\pm$ the internal $Spin(1,3)$ spinors, $\eta^\pm$, $\z^\pm$ the external $Spin(6)$ spinors of the Euclidean D5-brane. Unlike for the D3-brane, terms of both chirality need to be present to ensure reality, due to the fact that under conjugation spinors of $Spin(6)$ as well as those of $Spin(1,3)$ change chirality.
We normalize $\eta^+ = a \eta$ such that $\eta$ defines an $SU(3)$-structure via
\eq{
J_{mn} &= - i \we^c \g_{mn} \eta \\
\O_{mnp} &= \we \g_{mnp} \eta \;,
}
where the forms $(J, \O)$ satisfy
\eq{
\frac{1}{3!} J^3 = \left( \frac{-i}{2} \right)^3 \O \wedge \O^* = - \vol_6 \;.
}
We also set $\z^+ = b \eta$. Using this, we find that the non-linear instanton equations lead to
\eq{
\frac12 \fcal \wedge J \wedge J - \frac{1}{3!} \fcal \wedge \fcal \wedge \fcal &= \frac{b}{ai} \left( ( \sqrt{ \det( \delta + g^{-1}\fcal) } +1 ) \vol_6 + \frac12 \fcal \wedge \fcal \wedge J \right) \\
\left( \sqrt{ \det( \delta + g^{-1}\fcal) } -1 \right) \vol_6 - \frac12 \fcal \wedge \fcal \wedge J &= \frac{b}{ai} \left( \frac12 \fcal \wedge J \wedge J - \frac{1}{3!} \fcal \wedge \fcal \wedge \fcal\right) \\
\fcal \wedge \O &= 0 \;.
}
This is the six-dimensional analogue of \eqref{mmms1}, and leads to
\eq{\label{eq:mmms6d}
\frac12 \fcal \wedge J \wedge J - \frac{1}{3!} \fcal \wedge \fcal \wedge \fcal &= k \left(\vol_6 + \frac12 \fcal \wedge \fcal \wedge J\right) \\
\fcal^{(2,0)} &= 0 \;,
}
with $k = \frac{b}{ai} / (1 - (\frac{b}{ai})^2)$.  Once more, the formula matches the deformed instanton of  \cite{mmms} (see eq. (3.28)).
However, due to this exercise  we are able to make a clear distinction between the notions of linear and non-linear supersymmetry. Let us consider the linear case, which for us is defined as setting $\ve_- = 0$. Imposing this on \eqref{d=6nlinstanton} leads to the
following set of equations:
%, which are the proper six-dimensional analogue of the HYM condition \eqref{hym}:
\eq{\label{eq:sys6d}
\frac12 \fcal \wedge J \wedge J - \frac{1}{3!} \fcal \wedge \fcal \wedge \fcal &= 0 \\
\vol_6 + \frac12 \fcal \wedge \fcal \wedge J &= 0 \\
\fcal^{(2,0)} &= 0 \;.
}
As we see, in general the constrains imposed by linear supersymmetry appear to be more restrictive than those of  six-dimensional HYM equations. Like in $d=4$, due to the presence of $\rcal$ ``linear" supersymmetry yields instanton equation that are not linear in $\cal F$. However unlike $d=4$ case the two equations, defining the linear supersymmetry, are truly independent. Notice that taking $b=0$  ($k=0$) in the non-linear solution \eqref{eq:mmms6d} does not yield the full system and misses the first equation in  \eqref{eq:sys6d}.

As we shall discuss in Appendix \ref{sec:Lich}, one may find another justification for our definition of ``linear'' supersymmetry given here from the fact that by squaring it one finds a well-defined (albeit containing higher-deriative couplings)  action for six-dimensional Yang-Mills.

\section{Non-linear supersymmetries on curved spaces}\label{2d}
So far we have only discussed the linear and non-linear instanton equations originating for the D3-brane field theory. On a curved background these instantons can only be supersymmetric if the corresponding supersymmetry is preserved by the gravitational background. We will now discuss the supersymmetry conditions that govern the supersymmetric coupling of the D3-brane field theory to a curved background. This formalism will include both supersymmetries that are  linear and non-linear from the point of view of the worldvolume theory. In fact, we will find that the gravitational background does not differentiate between them. This is to be expected: the non-linear supersymmetries are associated to supersymmetries that are broken from the point of view of the worldvolume theory, but need not be broken from the point of view of the supergravity background. In the case with maximal supersymmetry, the D3-brane preserves 16 supercharges, whereas the gravity background preserves 32. Thus, we will see that the linear and non-linear R-symmetries of the D3-brane enhance to the full $SU^*(8)$ R-symmetry group of the supergravity background.
The coupling of field theories with less linear and non-linear supersymmetries to supergravity can be obtained by projecting out some of the supercharges from the $SU^*(8)$-symmetric theory. We will exemplify this by breaking the linear supersymmetry of the D3-brane down to $\ncal=1$ by using D7-branes in the background and obtain an $SU^*(2)$ R-symmetry group in the supergravity sector.
Let us expand on these ideas.

In \cite{fs}, it was shown how to construct $\ncal=1$ supersymmetric field theories on Riemannian manifolds. In particular, it was shown that in order to construct a field theory Lagrangian on a curved space, one should couple the field theory to a supersymmetric background of an off-shell $\ncal=1$ supergravity (in their case, use was made of both $\ncal=1$ ``old minimal" and ``new minimal" supergravity). Such supersymmetric backgrounds are determined by a pair of Killing spinor equations, which correspond to the vanishing of the gravitino variations for the specific $d=4$ off-shell supergravity theory to which the field theory is being coupled. This formalism can also be applied to Euclidean backgrounds by treating spinors of opposite chirality independently and taking the auxiliary fields to be complex.

In \cite{triendl}, it was shown that a pair of $d=4$ Killing spinor equations can be derived from type IIB supergravity by requiring compatibility of a supersymmetric fluxless Euclidean D3-brane with the external supersymmetry of the type IIB gravitino; in particular this was done by noting that
\eq{\label{hagen1}
\G_0 \ve &= \ve \\
D_m \ve &= 0 \;,
}
imply
\eq{\label{hagen2}
\{D_m, \G_0 \} \ve = 0 \;,
}
which can then be explicitly written in a manifestly four-dimensional way. Doing so, \eqref{hagen2} corresponds to the variation of the gravitino of $\ncal=4$ conformal supergravity.\footnote{See \cite{maxfield} for an investigation of $\ncal =4$ supergravity backgrounds to which $\ncal = 4$ SCFT can be coupled.} In order to break the supersymmetry down to $\ncal=1$, use was made of a pair of D7-branes. In fact, the $\ncal =1$ Killing spinor equations obtained in this way turn out to be more general than minimal supergravity, corresponding to 16/16 $\ncal =1$ supergravity \cite{16161,16162,16163}. The connection between the low-energy effective action of string theory and 16/16 supergravity (rather than minimal supergravity) was already made early on \cite{16162}.

The equation \eqref{hagen2} is a necessary but not a sufficient condition for the supersymmetry of the string theory background. In addition to the anti-commutator, it is necessary to consider also  the commutator. Furthermore, one should also consider the variations of the internal gravitino and the dilatino. The specific components of fluxes appearing in the commutator and the anti-commutator are different. In particular, the anti-commutator contains the internal rather than the external connection. Hence the equations arising from the anti-commutator may be considered as data specifying how to construct a full string background containing our four-dimensional supergravity background. Since the details of the field theory on the D3-brane are not affected, these can be ignored in our analysis. This is however not the case for the dilatino variation since the (modified) dilatino variations will lead to the vanishing of supersymmetry variations of other fermions in the background supergravity. For example, 16/16 supergravity can be considered as new minimal supergravity coupled to a chiral matter multiplet; the vanishing of the supersymmetry variation of the fermion of this multiplet can be obtained from the ten-dimensional dilatino variation.

In this section, we generalize this construction from the string theoretic point of view by allowing a Euclidean D3-brane to carry a non-trivial worldvolume flux. As we have shown in the previous section, the existence of non-linear instanton configurations of such a flux is related to the presence of non-linearly realized supersymmetries on the D3-brane. In other words, we will consider gravity backgrounds without imposing the first constraint of \eqref{hagen1}. Thus, D3-branes with a field theory with $\ncal$ linear supersymmetries couples to a background supergravity with $2 \ncal$ supersymmetries. We will explicitly demonstrate this for $\ncal =4$ instantons, coupling to $\ncal = 8$ supergravity, and $\ncal =1$ instantons, coupling to $\ncal =2$ supergravity.

In particular, rather than \eqref{hagen1}, our starting point will be
\eq{\label{genfs}
\G_0 \ve_\pm &= \pm \ve_\pm \\
D_m (\ve_+ + \ve_-) &= 0 \;.
}
The case of only linear supersymmetries can be obtained by setting the generator of the non-linear supersymmetries to zero, i.e.\ by taking $\ve_- = 0$.
For general supersymmetric configurations on the brane we will keep in mind that on the D3-brane with non-trivial worldvolume flux, the instanton equation \eqref{eq:nonlininstgen} holds in addition to \eqref{genfs}, and both equations have to be satisfied for the same supersymmetry parameter $\ve$, but otherwise both conditions are completely independent. Therefore we will focus now on the supersymmetry condition imposed by the gravitational background.

Let us consider how \eqref{genfs} generalizes \eqref{hagen1}. By not imposing the vanishing of $\ve_-$, we find that \eqref{hagen2} now has a non-trivial right-hand side, that is, we find
\eq{\label{ft}
\{D_m, \G_0\} \ve_+ &=  [D_m, \G_0 ] \ve_- \\
[D_m, \G_0 ] \ve_+ &= \{D_m, \G_0\} \ve_-  \;.
}
As we discussed, in the case of only linear supersymmetry, these two equations decouple. This is no longer the case, and one needs to solve both simultaneously in order to obtain the gravitino variations of the supergravity background.  In the rest of this section, we will do so explicitly. In a sense, this is the (Euclidean) IIB equivalent of the work done in \cite{deWit:1986mz}, except that we keep explicit track of which supersymmetries are linear and which are not. We will then demonstrate how the R-symmetry enhancement occurs. Furthermore in subsection \ref{sec:n=1}, we will discuss how the breaking to 1/4 supersymmetry works. A brief examination of  the dilatino variations of four-dimensional $\ncal =8$ and $\ncal =2$ supergravities will follow in subsection \ref{dilatinoapp}, and the link to the previous section on non-linear instantons is made in subsection \ref{ftconc}.

\subsection{Coupling $\ncal = 4$ field theory to an $\ncal = 8$ supergravity background }
We can now examine how the doublet equations \eqref{ft} reduce to dictate the supersymmetry conditions for $\ncal = 4$ four-dimensional Riemannian backgrounds. We use the terminology $\ncal=4$ to refer to the number of linear supersymmetries; the non-linear supersymmetries will double this amount. We consider the splitting of the $D=10$ Killing spinors $\ve_\pm$ into Killing spinors parallel and perpendicular to the D3-brane, and allow for both positive as well as negative chirality. Let us reiterate the specifics. We decompose $Spin(1,9) \rightarrow Spin(1,5) \otimes Spin(4)$ as
\eq{\label{ftksdecomp}
\ve_+^1 &= \xi^+_\a  \otimes \eta^+_\a +  \xi^-_\a  \otimes \eta^-_\a  \\
\ve_-^1 &= \xi^+_\a  \otimes  \z^+_\a  +  \xi^-_\a  \otimes  \z^-_\a \;,
}
where $\a \in \{1, ...,4\}$, since there are $16 = 4 \times 2 \times 2$  supercharges (with 2 components for both positive and negative chirality spinors in $d=4$). We will refer to the $Spin(1,5)$ spinors $\xi^\pm_\a$ as the ``internal spinor", and the $Spin(4)$ spinors $\eta^\pm_\a$ as the ``external spinors". From the four-dimensional point of view, the internal spinors can be viewed as the representation of the $\cN=4$ R-symmetry group; in the (externally) Lorentzian case, this is given by $Spin(6) \simeq SU(4)$, in the Euclidean case, we instead have $Spin(1,5) \simeq SU^*(4)$.
This is nothing more than a different choice of real form of the associated complexified Lie algebra.

The background supergravity does not distinguish between linear and non-linear realizations of supersymmetry on the D3-brane. Therefore we should combine the linear supersymmetries $\eta^\pm_\a$ and the non-linear supersymmetries $\z^\pm_\a$ into a spinor\footnote{We will suppress spinor-sum indices $\a$ when no confusion can arise.}
\eq{
\l^\pm = \left(\begin{array}{c} \eta^\pm \\ \z^\pm \end{array} \right) \;.
}
Since for the (Euclidean) background supergravity these supersymmetries are indistinguishable, we should find an $SU^*(8)$ symmetry that acts on $\l^\pm$ linearly, enhancing the $SU^*(4)$ symmetry of the linear supersymmetries.
Therefore the Killing spinor equations must admit an $SU^*(8)$ symmetry, and the components of the ten-dimensional bulk fields should reassemble in four-dimensional auxiliary fields in $SU^*(8)$ representations.
In the following we will show this explicitly by rewriting the Killing spinor equations originating from the external gravitino variation in an $SU^*(8)$-covariant form.

\subsubsection*{Killing spinor equations from the external gravitino}
We will eventually combine the equations \eqref{ft} into a pair of Killing spinor equations for $\l^\pm$. However, for symmetry purposes, it turns out that it will be convenient to use not just the supersymmetry variation of the ten-dimensional $\Psi_m$. We will instead consider\footnote{Note that the additional terms are spin-1/2 terms that can be removed by a superconformal transformation. Therefore the redefinition \eqref{extgravshift} can be ignored in the case of only coupling the linear supersymmetries to $\cN=4$ conformal supergravity, as has been done in \cite{triendl}.}
\eq{\label{extgravshift}
\delta \psi_m = \delta \Psi_m + \frac12 \g_m \delta \chi - \frac12 \g_m \G^a \delta \Psi_a \;.
}
The computation is rather lengthy, and is therefore relegated to appendix \ref{ftcalc}.\footnote{ There are two ways to perform the computation. Either one considers the modified gravitino from the start and then considers (anti-)commutators, or one first considers the (anti-)commutators of each term separately. We have done the latter.}
The final result that we find is that, making use of \eqref{ftksdecomp}, the ten-dimensional supersymmetry equations imply
\eq{\label{n=4ks}
\left( \nabla_m    + A_m^+   \right) \l^+ -\frac14 T_{np}^+ \g^{np} \g_m \l^- + K^+  \g_m \pcal_2 \l^- &\\
- \frac12 ( \p_m \phi- \frac12 \o^a_{\phantom{a}am}) \l^+ + \O_{mn}^+ \g^n \l^- + \Phi^+  \g_m  \pcal_2 \l^-   &= 0 \\
&\\
\left( \nabla_m    + A_m^-   \right) \l^- -\frac14 T_{np}^- \g^{np}  \g_m \l^+ + K^- \g_m \pcal_2 \l^+ &\\
- \frac12 ( \p_m \phi- \frac12 \o^a_{\phantom{a}am}) \l^- + \O_{mn}^- \g^n \l^+ +  \Phi^-  \g_m  \pcal_2 \l^+   &= 0 \;.
}
These equations are precisely the vanishing of the supersymmetry variations of the gravitino of $d=4$, $\ncal=8$ supergravity \cite{deWit:1986mz, NdW0, NdW2, wst}. No truncations have been performed, nor have any assumptions been made on the metric.\footnote{We have, however, assumed that $\ve_\pm$ can be chosen such that
\eq{
\nabla_m^{(4)} \xi_\a^\pm = \nabla_a^{(6)} \l^\pm_\a = 0\;.
}
}
We have chosen a gauge for the vielbeine such that
$e^\m_a = e^m_\a =0$, where $\m$ is a $4d$ and $\a$ a $6d$ curved index, such that $\o_{[ab]m} = 0$. We have written the equations in such a way that the first terms $A^\pm, T^\pm, K^\pm$ will have enhanced R-symmetry and furnish non-trivial $SU^*(8)$ representations, whereas the other terms do not; we will come back to this later.

Let us now give explicit expressions for all fields in \eqref{n=4ks}. They are defined in terms of the fluxes, as eigenvalues of the following Clifford action on the internal components of the Killing spinors. The composite connection $A_m^\pm$ is given by
\eq{\label{n=4fields1}
A_m^\pm  \left( \xi^\pm \l^\pm \right) =
\frac18  \cg^{ab} \xi^\pm \otimes
\left( \begin{array}{cc}
2 \o_{mab}  - \frac{1}{4!} i e^\phi \e_{abcdef} F_{m}^{\phantom{m}cdef} &
  H_{mab} \pm  i e^\phi F_{mab} \\
  H_{mab} \mp  i e^\phi F_{mab} &
2 \o_{mab}  + \frac{1}{4!} i e^\phi \e_{abcdef} F_{m}^{\phantom{m}cdef}
\end{array} \right)\l^\pm \\\
- \frac14 \xi^\pm \otimes
 \left( \begin{array}{cc}
\pm i e^{\phi} F_m &
\frac{1}{3!} \e_{mnpq} \left(\pm H^{npq} + i  e^\phi F^{npq} \right) \\
\frac{1}{3!} \e_{mnpq} \left(\pm H^{npq} - i  e^\phi F^{npq} \right)&
\mp i e^{\phi} F_m
\end{array} \right) \l^\pm \;.
}
Next, the field $K^\pm$ is given by
\eq{\label{n=4fields3}
K^{\pm } \left(\xi^\pm \l^\mp \right) = &
- \frac{1}{3!8} \cg^{abc} \xi^\mp \otimes
 \left( \begin{array}{cc}
\pm W_{abc} &
\pm H_{abc} +  i e^\phi F_{abc} \\
\pm H_{abc} -  i e^\phi F_{abc} &
\pm W_{abc}
\end{array} \right)  \l^\mp \\
&+ \frac18 \cg^a \xi^\mp \otimes
\left( \begin{array}{cc}
  i e^\phi \left(F_a \pm F_{a1234} \right) &
0 \\
0 &
- i e^\phi \left(F_a \pm F_{a1234} \right)
 \end{array} \right) \l^\mp \;.
}
Let us explain the terms $W_a, W_{abc}$. Group theory gives us the decomposition of the product representation ${ \bf 6} \otimes {\bf 4}$ of the vector and a chiral spinor of $SO(1,5)$. In a practical sense, this comes down to \cite{demedeiros}
\eq{\label{torsion}
\cg^a \nabla^{(6)}_a \xi^\pm &= W_a \cg^a \xi^\pm + \tilde{W}_{abc} \cg^a \cg^{bc} \xi^\pm \\
&=  W_a \cg^a \xi^\pm + \frac{1}{3!8} W_{a bc} \cg^{abc} \xi^\pm \;.
}
Here, $W_a$ and $W_{abc}$ can be expressed in terms of torsion classes and the (local) structure group defined by the Killing spinors $\xi^\pm_\a$. Note that there will certainly be constraints on $W_{abc}$, as a naive degree of freedom counting shows. In the Riemannian case, existence of covariantly contant spinors (i.e., $W_a = W_{abc}= 0$) requires the space to be flat. In the Lorentzian case, this is not necessary \cite{baum}.

We will generically assume that $W_a$ is exact; in this case, a Weyl rescaling can be found such that the term $\Phi^\pm$ defined by
\eq{
\Phi^\pm \left( \xi^\pm \l^\mp \right) =  \cg^a \xi^\mp \otimes \left(\p_a \phi - W_a\right) \l^\mp
}
can be trivialized.

Finally, we have the fields $T^\pm_{mn}$ given by
\eq{\label{n=4fields2}
T_{mn}^{\pm } \left(\xi^\pm \l^\mp \right) = &
\frac18 \cg^{abc} \xi^\mp \otimes
\left( \begin{array}{cc}
0 & \frac{1}{3!} i e^\phi F_{mnabc} \\
- \frac{1}{3!} i e^\phi F_{mnabc} & 0
\end{array} \right)  \l^\mp \\
&- \frac14  \cg^a \xi^\mp \otimes
\left( \begin{array}{cc}
\pm H_{mna} + i e^\phi F_{mna} &
\pm 2 \o_{amn}  \\
\pm 2 \o_{amn}  &
\pm H_{mna}  -  i e^\phi F_{mna}
 \end{array} \right) \l^\mp \;.
}
We have that $T_{mn}^+$ is anti-selfdual (and $T_{mn}^-$ is selfdual), which is also enforced by the contraction with the gamma matrices $\g^{np} \g_m$,
and hence satisfies
\footnote{We hope no confusion will arise from the superscript $\pm$, which refers to the fact that $T_{mn}^\pm$ is defined by acting on $\xi^\pm \l^\mp$ rather than to the (anti-)selfduality condition, which is exactly of the opposite sign.}
\eq{
- \frac14 T^\pm_{np}\g^{np}  \g_m \l^\mp = T^\pm_{mn} \g^n \l^\mp  \;.
}
This is in contrast to the term
\eq{
\O_{mn}^\pm \left(\xi^\pm \l^\mp \right) =
&  \mp \frac12  \cg^a \xi^\mp \otimes
\left( \begin{array}{cc}
0&
 \o_{amn} + \o_{mna}\\
 \o_{amn} + \o_{mna} &
0
\end{array} \right) \l^\mp \;,
}
which is not selfdual. These spin connection terms can be absorbed into the definition of $K^\pm$ instead, when working with curved indices, by noting that they can be written as a derivative acting on the curved vielbein. See \cite{deWit:1986mz} for details.

Finally, let us also note that we make use of a rescaled four-dimensional spin connection $\o_{mnp}$ in \eqref{n=4ks}, which differs from the usual spin connection $\ti{\o}_{mnp}$ by
\eq{\label{spinconshift}
\frac14 \o_{mnp} \g^{np}  = \frac14 \ti{\o}_{mnp} \g^{np} - \frac12 \g_{mn} (\p^n \phi -\frac12 \o_a^{\phantom{a}an}) \;.
}
In terms of the four-dimensional metric, the modification due to the dilatino comes about through the warping
\eq{
g_{\m\n}(x) = e^{- 2 \phi(x)} \ti{g}_{\m\n}(x) \;,
}
and provided that $\o_a^{\phantom{a}an}$ is exact, it can be similarly removed. This is the case for block diagonal warped metrics $g_{10} = g_4(x,y) + e^{2 \Delta(x)} g_6(y)$  for example.

Thus, under the assumption that we can do the Weyl rescaling as discussed above, we end up with
\eq{\label{n=4ks2}
\left( \nabla_m    + A_m^+   \right) \l^+ -\frac14 T_{np}^+ \g^{np} \g_m \l^- + K^+ \g_m \pcal_2 \l^-  &= 0 \\
\left( \nabla_m    + A_m^-   \right) \l^- -\frac14 T_{np}^- \g^{np} \g_m \l^+ + K^- \g_m \pcal_2 \l^+  &= 0 \;.
}
These equations govern the coupling of 16 linear and 16 non-linear supersymmetries to $\cN=8$ background supergravity. As we can see, background supergravity does not distinguish between linear and non-linear supersymmetries and treats them on equal footing. The auxiliary fields appearing in this equation are the $SU^*(8)$ connection $A_m^\pm$, selfdual and anti-selfdual two-tensors $T_{np}^\pm$ and a scalar $K^\pm$. We will now study their respective $SU^*(8)$ representations.

\subsubsection*{$SU^*(8)$ R-symmetry enhancement}
Let us comment on the representation theoretic interpretation of the fields appearing in \eqref{n=4ks2}. The R-symmetry group of the linear and non-linear supersymmetries, $SU^*(4) \times SU^*(4) \times \rbb^+$, enhances to the R-symmetry group of $\ncal=8$ supergravity, $SU^*(8)$. In the case, of a Lorentzian D3-brane, this would be the more familiar enhancement of $SU(4) \times SU(4) \times U(1) \rightarrow SU(8)$.

We note that the decomposition of the adjoint of $SU^*(8)$ to representations of $SU^*(4)  \times SU^*(4) \times \rbb^+$
is
\eq{
{\bf 63} \to ({\bf 15}, {\bf 1})_0 \oplus ({\bf 1}, {\bf 15})_0 \oplus ({\bf 4}, {\bf \bar 4})_{+2} \oplus ({\bf \bar 4}, {\bf 4})_{-2} \oplus ({\bf 1}, {\bf 1})_0 \ .
}
First, the diagonal subgroup of the two $SU^*(4)$ factors is just the geometric Lorentz group acting on the normal bundle, so it is generated by the $\cg^{ab}$, i.e.\
\begin{equation}
su^*(4)_{\rm diag} = \left\langle \cg^{ab} \otimes   \left( \begin{aligned} 1 && 0 \\ 0 && 1 \end{aligned} \right)\right\rangle \;,
\end{equation}
where $\langle T^{a b} \rangle$ denotes the span over $\rbb$ of generators $T^{ab}$.
The two commuting $su^*(4)$ algebras, which we will denote by $su^*(4)_\pm$, are then of the form
\begin{equation}
su^*(4)_\pm = \left\langle  \cg^{ab} \otimes \left( \begin{aligned} 1 && \pm 1 \\ \pm 1 && 1 \end{aligned} \right)\right\rangle \ .
\end{equation}
These are thus respectively the ${\bf(15,1)_0}$ and ${\bf (1,15)_0}$ representations.
The $\rbb^+$ that commutes with them, i.e., the ${\bf(1,1)_0}$, is generated by
\begin{equation}\label{u1_su8}
\rbb = \left\langle  \cg_{(6)} \otimes \left( \begin{aligned} 0 && 1 \\ 1 && 0 \end{aligned} \right)  \right\rangle \ .
\end{equation}

Now let us identify the remaining $({\bf 4}, {\bf \bar 4})_{+2}$ and $({\bf \bar 4}, {\bf 4})_{-2}$ representations. We want some representations where the two $SU^*(4)$ factors only act from the right or left, but in total these representations need two $SU^*(4)$ indices. The matrices $\{\cg_{ab}, 1\}$ form a basis of four-by-four matrices on which $\cg_{ab}$ can act from the left as a ${\bf \bar 4}$ or from the right as a ${\bf 4}$. The two-by-two matrices then have to ensure that $su^*(4)_\pm$ only acts from the right (left) on $({\bf 4}, {\bf \bar 4})_{+2}$, and from the left (right) on $({\bf \bar 4}, {\bf 4})_{-2}$. Last but not least both representations should have definite $\rbb^+$ charges, which means that their pair of off-diagonal two-by-two components and their pair of diagonal components each must be anti-symmetric.
We find that
\begin{equation}
({\bf 4}, {\bf \bar 4})_{+2} = \left\langle
 \cg^{ab} \otimes\left(  \begin{array}{cc} 1  &  -\cg_{(6)} \\ \cg_{(6)} & - 1 \end{array} \right) ,~
  1       \otimes\left(  \begin{array}{cc} \cg_{(6)} &  - 1 \\ 1 & - \cg_{(6)} \end{array} \right)
 \right\rangle
\end{equation}
and
\begin{equation}
({\bf \bar 4}, {\bf 4})_{-2} = \left\langle
\cg^{ab}\otimes \left( \begin{array}{cc} 1 &  \cg_{(6)} \\ - \cg_{(6)} & - 1 \end{array} \right) ,~
 1      \otimes \left( \begin{array}{cc} \cg_{(6)} &   1 \\ - 1 & - \cg_{(6)} \end{array} \right)
\right\rangle
\end{equation}
are a basis of generators that do the job.

We will also need the ${\bf 28}$ and the ${\bf 36}$ representations of $SU^*(8)$, as well as their complex conjugates:\footnote{Specifically, these would be the primitive $(2,0)$-form and the primitive $(1,7)$-form in the $SU(8)$ case.}
\eq{
\begin{alignedat}{6}
{\bf 28}       &\to ({\bf 6}, {\bf 1})_{+2}  \oplus ({\bf 1}, {\bf 6})_{-2}  \oplus ({\bf 4}, {\bf 4})_{0}  && ~~ ,\qquad &&
{\bf \bar{28}} &&\to ({\bf 6}, {\bf 1})_{-2}  \oplus ({\bf 1}, {\bf 6})_{+2}  \oplus ({\bf \bar{4}}, {\bf \bar{4}})_{0} \\
{\bf 36}       &\to ({\bf 10}, {\bf 1})_{+2} \oplus ({\bf 1}, {\bf 10})_{-2} \oplus ({\bf 4}, {\bf 4})_{0}  && ~~ , \qquad &&
{\bf \bar{36}} &&\to ({\bf 10}, {\bf 1})_{-2} \oplus ({\bf 1}, {\bf 10})_{+2} \oplus ({\bf \bar{4}}, {\bf \bar{4}})_{0} \;.
\end{alignedat}
}
The ${\bf (6,1)}$ and ${\bf (1,6)}$ representations can be determined by the following constraints: their $\rbb^+$-charge, the fact that they commute with one of the $su^*(4)$-factors, and the number of degrees of freedom. We thus find
\eq{
({\bf 6}, {\bf 1})_{\pm 2} &=\left\langle \pm \frac12 \left(1 \pm \cg_{(6)}\right) \cg^{a} \otimes \left(\begin{aligned} 1 && 1 \\ 1 && 1 \end{aligned} \right) \right\rangle \\
({\bf 1}, {\bf 6})_{\mp 2} &=\left\langle \pm \frac12 \left(1 \pm \cg_{(6)}\right) \cg^{a} \otimes\left(\begin{aligned} 1 && -1 \\-1 && 1 \end{aligned} \right) \right\rangle
\;.
}
Similarly, we also construct the ${\bf (10,1)}$ and the ${\bf (1,10) }$ representations:
\eq{
\begin{alignedat}{6}
({\bf 10}, {\bf 1})_{\pm 2} &=
\left\langle \pm \frac{1}{3!2}\left(1 \pm \cg_{(6)}\right) \cg^{abc} \otimes \left(\begin{aligned} 1 && 1 \\ 1 &&1 \end{aligned} \right) \right\rangle
\\
({\bf 1}, {\bf 10})_{\mp 2} &=
\left\langle \pm \frac{1}{3!2} \left(1 \pm \cg_{(6)}\right) \cg^{abc} \otimes \left(\begin{aligned} 1 && -1 \\-1 && 1 \end{aligned} \right)\right\rangle
\;.
\end{alignedat}
}
Finally, by noting that the algebra should close, we construct the representations
\eq{ \label{eq:bispinor_rep}
({\bf 4}, {\bf 4})_{0} &=
\left\langle  \frac12 \left(1 + \cg_{(6)}\right) \cg^{a} \otimes \left( \begin{aligned} 1 &&  0 \\ 0 && - 1 \end{aligned} \right) ,
~\frac{1}{3!2} \left(1 + \cg_{(6)}\right)  \cg^{abc}  \otimes\left( \begin{aligned} 0 &&  1 \\ -1 && 0 \end{aligned} \right) \right\rangle \\
({\bf \bar{4}}, {\bf \bar{4}})_{0} &=
\left\langle  \frac12 \left(1 - \cg_{(6)}\right)\cg^{a} \otimes \left( \begin{aligned} 1 &&  0 \\ 0 && - 1 \end{aligned} \right),
~\frac{1}{3!2} \left(1 - \cg_{(6)}\right) \cg^{abc} \otimes \left( \begin{aligned} 0 &&  1 \\ -1 && 0 \end{aligned} \right)  \right\rangle \;.
}
%{\bf [DP: Also, hermiticity properties need to be checked for all generators.]}.\\
These form bispinors under $SU^*(4)_+ \times SU^*(4)_-$, and the two components of the representation correspond to the symmetric and anti-symmetric parts.\footnote{In order to see this, compare the action of $SU^*(4)_\pm$ onto \eqref{eq:bispinor_rep} to the left and right action of $SU^*(4)$ onto $\{ \cg^a, \cg^{abc}\}$.}

Now let us apply this group theory to the Killing spinor equations \eqref{n=4ks2} that we found. Our claim is that
\eq{
\begin{alignedat}{5}
A_m^\pm  &\sim {\bf 63} &&&&&&\\
T_{mn}^+ &\sim {\bf 28}  && \qquad && T_{mn}^- &&\sim \bar{ {\bf28} }\\
K^+ &\sim {\bf 36} && \qquad && K^- &&\sim \bar {\bf 36}\;.
\end{alignedat}
}
First, we spell out the field $A_m^\pm$, given by
\eqref{n=4fields1}, explicitly in terms of the generators we have written above. $A_m^\pm$ is the composite connection, and should therefore be the adjoint, i.e.\ the  $\bf {63}$ of $SU^*(8)$. Let us generically denote generators of a representation ${\bf (k,l)}$ as $T_{\bf (k,l)}$. We then have that (suppresing representation indices)
\eq{\label{n=4rep1}
A^\pm_m  = \frac18 \Big[&
                    \o_{mab}                                  \left( T_{\bf(15,1)}^{ab} + T_{\bf (1,15)}^{ab} \right)
 + \frac12           H_{mab}                                  \left( T_{\bf (15,1)_0}^{ab} - T_{\bf (1,15)_0}^{ab} \right)\\
&- \frac{1}{4!2} i e^\phi \e_{abcdef} F_{m}^{\phantom{m}cdef} \left(T_{\bf (\bar{4}, 4)_{-2}}^{ab} + T_{\bf (4,\bar{4})_{+2}}^{ab} \right)
 + \frac12       i e^\phi F_{mab}                             \left(T_{\bf (\bar{4}, 4)_{-2}}^{ab} - T_{\bf (4,\bar{4})_{+2}}^{ab} \right) \\
&-               i e^\phi F_m                                 \left(T_{\bf (\bar{4}, 4)_{-2}} + T_{\bf (4,\bar{4})_{+2}} \right)
 - \frac{1}{3!}  i e^\phi \e_{mnpq} F^{npq}                   \left(T_{\bf (\bar{4}, 4)_{-2}} - T_{\bf (4,\bar{4})_{+2}} \right) \\
&- \frac{2}{3!}           \e_{mnpq} H^{npq} ~                       T_{\bf (1,1)_0}
\Big] \;.
}
We see that the representation decomposition is indeed exactly that of the ${\bf 63}$.

Let us repeat the process for the field $T^\pm_{mn}$, given in \eqref{n=4fields2}.\footnote{We hope no confusion will arise between the supergravity field $T^\pm_{mn}$ and the generators of the Lie algebras labeled by $T_{\bf(k,l)}$.} We see that
\eq{\label{n=4rep2}
T_{mn}^+ \left( \xi^+ \l^-\right) =  \frac18 \Big[&
             i e^\phi   F_{mnabc}        T_{{\bf(4,4)_{0}}}^{abc}
  -       2  i e^\phi   F_{mna}          T_{{\bf(4,4)_{0}}}^a \\
& -                     H_{mna}   \left( T_{{\bf(6,1)_{+2}}}^a + T_{{\bf(1,6)_{-2}}}^a \right)
  -       2            \o_{amn}   \left( T_{{\bf(6,1)_{+2}}}^a - T_{{\bf(1,6)_{-2}}}^a \right)
  \Big]  \xi^- \l^- \;,
}
with anti-selfduality implicit. We thus see that $T_{mn}^+$ is given exactly by the ${\bf 28}$ of $su^*(8)$.\footnote{In fact, note that if one prefers, one might as well add the $\bar{{\bf 28}}$, which drops outs due to the chiral projection operators in the generators of the representation.} As expected, we find that
$T_{mn}^-$ fills out the ${\bf \bar{28}}$:
\eq{
T_{mn}^- \left(\xi^- \l^+ \right) =   \frac18 \Big[&
                i e^\phi F_{mnabc}      T_{{\bf(\bar{4},\bar{4})_{0}}}^{abc}
 -       2      i e^\phi F_{mna}        T_{{\bf(\bar{4},\bar{4})_{0}}}^a\\
&-                       H_{mna} \left( T_{{\bf(6,1)_{-2}}}^a + T_{{\bf(1,6)_{+2}}}^a \right)
 -       2              \o_{amn} \left( T_{{\bf(6,1)_{-2}}}^a - T_{{\bf(1,6)_{+2}}}^a \right)
  \Big] \xi^+ \l^+ \;.
}
Again, selfduality is implied.

Finally, there is the field that comprises the conformal transformation, $K^\pm$, which decomposes as
\eq{
K^+ \left(\xi^+ \l^-\right) = - \frac{1}{16} \Big[ &
                   H_{abc}              \left( T_{\bf(10,1)_{+2}}^{abc} - T_{\bf (1,10)_{-2}}^{abc} \right)
+                  W_{abc}              \left( T_{\bf(10,1)_{+2}}^{abc} + T_{\bf (1,10)_{-2}}^{abc} \right) \\&
+ 2 i e^\phi       F_{abc}                     T_{\bf (4,4)_0}^{abc}
- 2 i e^\phi \left(F_a + F_{a1234} \right)     T_{\bf (4,4)_0} \Big] \xi^- \l^+ \;,
}
and
\eq{
K^-\left( \xi^- \l^+\right) = - \frac{1}{16} \Big[ &
                   H_{abc}                \left( T_{\bf(10,1)_{-2}}^{abc} - T_{\bf (1,10)_{+2}}^{abc} \right)
+                  W_{abc}                \left( T_{\bf(10,1)_{-2}}^{abc} + T_{\bf (1,10)_{+2}}^{abc} \right) \\&
+ 2 i e^\phi       F_{abc}                       T_{\bf {(\bar{4},\bar{4})}_0}^{abc}
- 2 i e^\phi \left(F_a - F_{a1234} \right)       T_{\bf {(\bar{4},\bar{4})}_0} \Big] \xi^+ \l^+ \;.
}
As was remarked before, we see that these representations are respectively the ${\bf 36}$ and the ${\bar {\bf 36}}$.

\subsection{Coupling $\ncal =1$ field theory to an $\ncal=2$ supergravity background}\label{sec:n=1}

In the previous section we worked out the coupling of $\ncal =4$ field theory with an additional non-linear $\ncal =4$ supersymmetry to background $\ncal =8$ supergravity. In general also field theories with less linear and non-linear supersymmetries can be coupled to background supergravity.
To illustrate this, let us now reduce the supersymmetric field theory and its coupling to supergravity of the last section to an $d=4$ Euclidean supersymmetric field theory with $\ncal = 1$ linear supersymmetry and an additional $\ncal = 1$ non-linear supersymmetry.

In order to break the supersymmetry $\ncal = 4 \rightarrow \ncal=1$, we will introduce a pair of D7-branes that fill out the entire D3-brane. The logic here is that the kappa-symmetry requirements of the D7-branes project out a number of degrees of freedom of the Killing spinor $\ve$, such that, if the projection operators are chosen appropriately, one is left with exactly 1/4 of the orginal supercharges, i.e.\ four linear supercharges when starting with sixteen. A priori, it is not clear what the properties of the D7-branes should be to obtain the appropriate projections. There appear to be two candidates more natural than any other choice. In both of these cases that we will describe, both D7-branes are Lorentzian, one of them localized in the $X^{8,9}$ directions, the other localized in the $X^{6,7}$ directions. The difference between the two cases lies in the worldvolume flux $\fcal$. We consider the following possibilities:
\begin{itemize}
\item Both D7-branes carry no worldvolume flux.
\item Both D7-branes carry worldvolume flux, which is the pullback of the worldvolume flux of the Euclidean D3-brane.
\end{itemize}
It turns out that the compatibility of the first option with the supersymmetry of non-trivial worldvolume flux on the D3-brane is problematic, as we demonstrate in appendix \ref{d7}, so let us discuss the second option instead.

We denote the kappa-symmetry operators of the D3-brane and the D7-branes with worldvolume flux as respectively $\G^3$, $\G^7$, $\G^{\ti{7}}$. Then supersymmetry of all three branes is equivalent to enforcing
\eq{
\G^7 \ve= \G^{\ti{7}} \ve = \G^3 \ve = \ve \;.
}
Taking into account our ansatze for the worldvolume flux, the kappa-symmetry operator for the Lorentzian D7 branes is given by
\eq{
\G^7 &= - i \g_{0567} \G^3 \\
\G^{\ti{7}} &= - i \g_{0589} \G^3 \;.
}
Using this and the fact that $\G_{(10)} \ve \equiv \g_{(4)} \cg_{(6)} \ve = \ve$, it follows that
\eq{\label{g2chi2}
\cg_{05} \xi^\pm &= \pm   \xi^\pm\\
\cg_{67} \xi^\pm &= \pm i \xi^\pm\\
\cg_{89} \xi^\pm &= \pm i \xi^\pm \;.
}
This leads to the conclusion that
\eq{
\xi^- = \cg_{068} \xi^+ \;,
}
up to a constant prefactor which we trivialize without loss of generality.
We will now proceed to obtain the $\ncal = 1$ equations. The derivation will be somewhat messy and will therefore be relegated to appendix \ref{ftcalc}. In essence, we will make use of \eqref{g2chi2} to construct certain projection operators, and then figure out with which operators we should act on
\eqref{n=4ks2} to obtain the desired Killing spinor equations. In the case $\ve_- = 0$, this procedure leads to
\eq{\label{n=1lfs}
\left( \nabla_m + i A_m + i V_n \g^n \g_m  \right) \eta^+ - M^+ \g_m \eta^- &= 0\\
\left( \nabla_m - i A_m - i V_n \g^n \g_m  \right) \eta^- - M^- \g_m \eta^+ &= 0 \;,
}
which are the gravitino variations of 16/16 $\ncal=1$ supergravity. More generally, the presence of $\ve_-$ will lead to the gravitino variations of an $\ncal = 2$ supergravity (see for example \cite{andrianopoli}), where we have the embedding of the $\ncal=1$ case spelled out explicitly.

The projection operators break the $\ncal=4$ R-symmetry group down to the $\ncal=1$ R-symmetry group, which then enhances to the $\ncal=2$ R-symmetry group for non-trivial $\ve_-$. In our case, where we examine Euclidean backgrounds for field theories, that comes down a breaking of $SU^*(4) \times SU^*(4) \times \rbb^+$ down to $\rbb^+ \times \rbb^+$, which then enhances to $SU^*(2) \times \rbb^+$. In the more familiar case of Lorentzian field theory, the analogue would be breaking $SU(4) \times SU(4) \times U(1)$ to $U(1) \times U(1)$, which then enhances to $U(2)$.
Specifically, the $SU(4)$ acting on the normal bundle would be broken by the projection to $U(1)^3$, generated by $\cg_{45}$, $\cg_{67}$ and $\cg_{89}$. Let us discuss the adjoint first. The two $SU^*(4)_\pm$ factors are projected to
\begin{equation}
(\rbb)_\pm = \left\langle \left( \begin{array}{cc} 1 & \pm 1 \\ \pm 1 & 1 \end{array} \right)\otimes \cg_{(6)} \right\rangle \ .
\end{equation}
Note that the $\rbb^+$ generated by \eqref{u1_su8} is a subgroup of these two factors.
Furthermore, the $({\bf 4}, {\bf \bar 4})_{+2}$ and $({\bf \bar 4}, {\bf 4})_{-2}$ are projected to
\begin{equation}
\left\langle \left( \begin{array}{cc} \cg_{(6)} & -1 \\ 1 & - \cg_{(6)} \end{array} \right), ~~~\left( \begin{array}{cc} \cg_{(6)} &  1 \\ -1 & - \cg_{(6)} \end{array} \right) \right\rangle\ .
\end{equation}
In total this gives in the Lorentzian case the gauge group $SU(2) \times U(1)$, as the eigenvalues of $\cg_{(6)}$ are imaginary. In the case of a Euclidean field theory, the eigenvalue of $\cg_{05}$ differs by a factor of $i$ from the one for $\cg_{67}$ and $\cg_{89}$, and so the gauge group is complexified. Furthermore, $\cg_{(6)}$ has real eigenvalues and so the gauge group is $SU^*(2) \times \mathbb{R}^+$.

The Killing spinor equations we obtain by the above procedure are given by
\eq{\label{n=1fs}
\left( \nabla_m - \frac12 \p_m \phi + i A_m^{0} + i A_m^{j} \s_j   \right) \l^+ - \frac14  T_{np}^+ \g^{np} \g_m  \e^{(2)} \l^- + \g_m M^{+j} \s_j   \e^{(2)} \l^-  &= 0 \\
\left( \nabla_m - \frac12 \p_m \phi - i A_m^{0} - i A_m^{j} \s_j^* \right) \l^-  -\frac14  T_{np}^- \g^{np} \g_m  \e^{(2)} \l^+ + \g_m M^{-j} \s_j^* \e^{(2)} \l^+  &= 0 \;.
}
Here, $\s_j$ are the Pauli-matrices generating $SU^*(2)$ and
\eq{
\e^{(2)} \l^\pm = \left(\begin{array}{cc} 0&1 \\-1&0 \end{array} \right) \l^\pm \;.
}
The $d=4$ fields are expressed in terms of the string theory fields as
\eq{\label{n=2fields1}
A^0_m  &=   \frac12 (i \o_{m05} +  \o_{m67} +  \o_{m89})  \\
A^1_m  &=   \frac14           (i H_{m05} + H_{m67} + H_{m89} + \frac{1}{3!} i \e_{mnpq} H^{npq} )  \\
A^2_m  &= - \frac14  e^\phi   (i F_{m05} + F_{m67} + F_{m89} + \frac{1}{3!} i \e_{mnpq} F^{npq} )  \\
A^3_m  &= - \frac14  e^\phi   (F_m - F_{m6789} - i F_{m0567} - i F_{m0589}) \ ,
}
and
\eq{\label{n=2fields2}
T_{mn}^\pm &= - \frac18 i \e^{\phi} \Big( F_{mn068} - F_{mn079} + i (F_{mn569} + F_{mn578} ) \\
& \phantom{= \frac{1}{8} \e^{\phi} \Big(} \mp F_{mn568} - F_{mn579} + i (F_{mn069} + F_{mn078} ) \Big) \;,
}
and
\eq{\label{n=2fields3}
M^{\pm1} & = -  \frac18 i e^\phi \Big( F_{068} - F_{079} + i F_{569} + i F_{578}  \mp \left(F_{568} - F_{579} + i F_{069} + i F_{078} \right)   \Big) \\
M^{\pm2} & =    \frac18 i        \Big( H_{068} - H_{079} + i H_{569} + i H_{578}  \mp \left(H_{568} - H_{579} + i H_{069} + i H_{078} \right)   \Big) \\
M^{\pm3} & =\pm \frac18          \Big( W_{068} - W_{079} + i W_{569} + i W_{578}  \mp \left(W_{568} - W_{579} + i W_{069} + i W_{078} \right)   \Big) \\
}
Finally, the spin connection in is shifted by the dilatino, exactly as given in \eqref{spinconshift}. This matches precisely the variation of the gravitino in $d=4$, $\ncal =2$ supergravity, as given in \cite{andrianopoli} (see eq. (8.24)). All terms exhibit manifest R-symmetry invariance.

\subsection{The dilatino variation}\label{dilatinoapp}
We have discussed so far how to obtain the gravitino variations of four-dimensional $\ncal =8$ and $\ncal =2$ supergravities  from the supersymmetry conditions of type IIB supergravity and of D3-branes. In order to fully determine supersymmetric backgrounds for field theories, the variations of all fermions of the supergravity that the field theory is being coupled to need to vanish. Both $\ncal =8$ and $\ncal =2$ contain, in addition to the gravitino, another fermion, which we will refer to here as the ``dilatino". In the case of (Lorentzian) maximal $\ncal = 8$ supergravity, the dilatino $\hat{\chi}_{\hat{A}\hat{B}\hat{C}}$ transforms under the ${\bf 56}$ of $SU(8)$, and the vanishing of its variation corresponds to (compare with \cite{deWit:1986mz, NdW0, NdW2, wst})
\eq{\label{dilatinovar}
- \frac34 \sqrt{2} T^+_{mn[\hat{A}\hat{B}}  \g^{mn} \l^+_{\hat{C}]} + P^+_{m\hat{A}\hat{B}\hat{C}\hat{D}}   \g^m \l^{- \hat{D}}  + ... &= 0 \\
- \frac34 \sqrt{2} T^{-[\hat{A}\hat{B}}_{mn} \g^{mn} \l^{-\hat{C}]} + P_{m}^{-\hat{A}\hat{B}\hat{C}\hat{D}} \g^m \l^+_{\hat{D}}  + ... &= 0 \;,
}
with the ellipsis representing terms that correspond to Weyl transformations and with $\hat{A}, \hat{B}, .. \in \{1, ...,8\}$ being fundamental $SU(8)$ indices. Here, $T^\pm$ is the same field as appearing in the gravitino variation. Thus, in order to complete the derivation of the constraints on four-dimensional backgrounds from a string theory perspective, it would be necessary to understand how to obtain the above equations, with the four-dimensional field $T^\pm$ matching \eqref{n=4fields2}
and $P^\pm$ having some given expression in terms of the string fields.

From an M-theory point of view, this can be achieved by setting
\eq{
\hat{\chi}_{\hat{A}\hat{B}\hat{C}} \sim \left( 1 + \g_{(4)} \right) \G^a_{[ \hat{A} \hat{B} } \Psi_{\hat{C}]a}\;,
}
with the $Spin(7)$ indices being lifted to $SU(8)$. The naive analogous terms in IIB would be the building blocks
\eq{
\left(1 \pm \G_{(10)} \cdot \right) \G^a_{[AB} \Psi_{C]a} \\
\left(1 \pm \G_{(10)} \cdot \right) \G_{a[AB} \G^a_{C]D} \chi^D \;,
}
where the indices $A, B, ...$ are $SU(4)$ indices. Taking a linear combination of the two and proceeding as before for the gravitino (as outlined in section \ref{ftcalc}), one is able to arrive at a variation of the same general form as \eqref{dilatinovar}, but with coefficients of the string fields such that one cannot quite reconstruct $T^\pm$.

We believe the reason for this is as follows. Under the decomposition $SU(8) \rightarrow SU(4) \times SU(4) \times U(1)$, an $SU(8)$ index $\hat{A}$ corresponds to an $SU(4)$ index tensored with a doublet structure; for example, this can be observed in the definition of $T^\pm_{mn\hat{A}\hat{B}}$ in \eqref{n=4fields2}: roughly speaking, the two $SU(8)$ indices correspond to the two $SU(4)$ indices (i.e., the indices of the gamma-matrices) tensored with a two-tensor in doublet space (i.e., the $ 2 \times 2$ matrix acting on $\l^\pm$). Therefore, in order to construct (the variation of) $\hat{\chi}_{\hat{A}\hat{B}\hat{C}}$, one should construct an object with three $SU(4)$ indices which is a three-tensor in doublet space. However, our methods fundamentally rely on acting on the doublet structure with $\G_0$: such matrix multiplication cannot produce a three-tensor. Thus, one requires a different recipe for obtaining the desired dilatino variation of maximal four-dimensional supergravity. While completing such a construction may be interesting for producing a full IIB version of \cite{deWit:1986mz}, this stays outside the scope of our paper.

In the case of $\ncal = 2$ supergravity, the dilatino does not suffer from this issue, and so in theory one should be able to construct it using our methods. However, the problem for $\ncal =2$ supergravity is that, since the fields appearing in the variation of the dilatino and those appearing in the variation of the gravitino are distinct, there is no way to compare which ten-dimensional object leads to the ``right" dilatino: any spinor transforming as the fundamental of the R-symmetry group would do.

\subsection{Instantons on curved backgrounds}\label{ftconc}

We have shown how to reformulate the bulk supersymmetry conditions in a way that describes the coupling of the supersymmetric field theory on the brane world volume to background supergravity. The equations in \eqref{n=4ks} and their simplified version \eqref{n=4ks2} give an $SU^*(8)$-covariant description those couplings. In particular we see that background supergravity couples universally to all supersymmetries and does not distinguish between linear and non-linear realizations on the worldvolume. Hence non-linear supersymmetries can be coupled to background supergravity in exactly the same way as linear supersymmetries. Only when embedded in a ten-dimensional string background, these theories couple to different bulk fields.

We can obtain theories with less linear and non-linear supersymmetries by projecting out some of the supersymmetries. We found for instance a theory with four linear and four non-linear supersymmetries the supersymmetry condition \eqref{n=1fs} for its coupling to background supergravity. In general also theories with different amounts of linear and non-linear supersymmetries can be studied.

In summary, the supersymmetry conditions for theories with linear and non-linear supersymmetries are given by the worldvolume supersymmetry condition \eqref{eq:nonlininst}, which can be interpreted as the gaugino variation of the field theory, and the condition for a supersymmetric coupling to background supergravity \eqref{n=4ks} or \eqref{n=4ks2}.

Combining both supersymmetry conditions, we can study general instanton configurations for a supersymmetric field theory on a curved background.  For instance,  configurations such that only linear combinations of linear and non-linear supersymmetries are coupled supersymmetrically to the gravitational background fields can be engineered. This  should lead to supersymmetric field theories on curved backgrounds with a non-supersymmetric vacuum and no linear instanton solutions, but with supersymmetric non-linear instanton solutions. It would be interesting to study the possibilities for localization techniques for such backgrounds.

\section{Conclusion}

In this work we discussed non-linear instantons of supersymmetric field theories on curved backgrounds that originate from D3-brane theories in type IIB. We showed that the appearance of non-linear instantons is  tied to  existence of the supersymmetries that are spontaneously broken in the vacuum of the field theory on flat space and  are therefore non-linearly realized in the worldvolume theory.

In order to put the field theory on a curved space in a supersymmetric fashion, one needs to ensure that the coupling to the background fields does not violate these non-linear supersymmetries. In other words, the supergravity background to which the field theory in question is coupled is also supersymmetric with respect to non-linear supersymmetries. As we showed here, the ten-dimensional supersymmetry conditions can be used to express the supersymmetry variations of the fermions in that off-shell supergravity multiplet. In particular, the supersymmetries which are non-linearly realized on the worldvolume, are linearly realized on the supergravity background.  The supergravity background has effectively double the amount of supercharges of the field theory. The supersymmetry condition coming from the external gravitino variation can be written in a covariant form, by modifying it by the dilatino and the trace of the internal gravitino variation. Similarly we suspect the dilatino variation and the internal gravitino variation to give rise to additional supersymmetry conditions.
An overview of our discussion is found in Table~\ref{table}.

\begin{center}
\begin{table}[h]
\begin{tabular}{ |l |l|l|l|l|} \hline
 \backslashbox{SUGRA}{FT} & \cellbreak{bulk\\ geometry} & \cellbreak{FT from D$3$:\\ $\G_0 \ve = \ve $} &\cellbreak{only linear SUSY:\\$\G_0 \ve= \ve$, $\G \ve= \ve$} & \cellbreak{general SUSY:\\ $\G \ve= \ve$} \\ \hline
\cellbreak{pure geometry:\\ $\nabla_m \ve= 0$}             & \cellbreak{special\\ holonomy} & \cellbreak{topologically\\ twisted FT}     & \cellbreak{Hermitian\\ Yang-Mills (HYM)}           & \cellbreak{non-linear \\ instanton (NLI)}  \\ \hline
\cellbreak{with fluxes:\\ $ D_m \ve=  0$}              & \cellbreak{generalized\\ geometry} & \cellbreak{FT  coupled to\\  off-shell SUGRA}  & \cellbreak{HYM coupled to\\  off-shell SUGRA}          &  \cellbreak{NLI coupled to\\  off-shell SUGRA} \\ \hline
\end{tabular}
    \caption{A sketch of the different cases of a field theory (FT) originating from a D3-brane in a given supergravity (SUGRA) background.}
    \label{table}
\end{table}
\end{center}

While the paper is largely devoted to D3-branes and four-dimensional theories, we also discussed supersymmetry variations and instantons for a D5-brane. Here even the supersymmetries that are unbroken in the vacuum have a non-linear dependence on the worldvolume flux $\fcal$ in the variation of the fermions. Consequently, supersymmetric instantons in six dimensions are naturally non-linear.

Naturally one would like to find examples of non-linear instanton configurations in string theory. A D3-brane with non-linear instanton configuration will in general not preserve the same supersymmetries that are unbroken in standard warped Calabi-Yau orientifold compactifications of type IIB. Instead these D3-brane configurations are found in more general flux backgrounds, compatible with the projection \eqref{eq:kappa_symm}. Examples of such backgrounds can for instance be found along the baryonic branch connecting Klebanov-Strassler and Maldacena-Nunez backgrounds \cite{baryonic}.

The non-linearly realized supersymmetries and the related instantons might be of interest in various applications for supersymmetric field theories. So far localization computations have only used linearly realized supersymmetries. In principle non-linearly realized supersymmetries, leading to the theory that localizes on the non-linear instantons discussed above, can also be used for localization. In the case where also linearly realized supersymmetries are present the computation should be independent of which supercharge is used for localization, which  should lead to a non-trivial correspondence between linear and non-linear instantons. Localization on non-linear instantons can however also be used in field theories where supersymmetry is spontaneously broken in the vacuum, as in that case all supersymmetries are non-linearly realized. This in turn may lead to  applications of supersymmetric localization for non-supersymmetric field theories.

A particular example of a non-supersymmetric field theory with non-linearly realized supersymmetries is give by  an anti-D3 brane in a Calabi-Yau background with imaginary selfdual three-form flux, as discussed in \cite{gkp, gp,  kpv}. All linear supersymmetries of the worldvolume theory are explicitly broken by the supergravity background in that case. However  the coupling to the supergravity background preserves the non-linear supersymmetries.  Hence we may hope that the better control of non-linearly realized supersymmerty and the methods highlighted in this work might lead to better understanding of the field theories on anti-branes in flux backgrounds.

%%%%%%%%%%%%%%%%%%%%%%%%%%
\section*{Acknowledgments}
%%%%%%%%%%%%%%%%%%%%%%%%%%
We thank Thomas Dumitrescu, Charlie Strickland-Constable, Valentin Reys and Dan Waldram for useful discussions.
R.M and D.P. would like to thank KIAS for hospitality during the course of this work.
This work was supported in part by the Agence Nationale de la Recherche under the grant 12-BS05-003-01 (R.M.), by ERC Grant Agreement n. 307286 (XD-STRING) (D.P.) and by EPSRC Programme Grant EP/K034456/1 (H.T.).

\clearpage

\bookmarksetupnext{level=part}
\begin{appendices}

\section{Conventions and identities}\label{sec:C&I}
The connection is given by
\eq{
\nabla_M = \p_M + \frac14 \o_{MNP} \G^{NP}\;.
}
The sign we use for the Lorentzian Levi-Civita is
\eq{
\epsilon_{m_1 ...} = - \e^{m_1 ...} = + 1\;.
}
Our definition for the Hodge-dual of a a $k$-form $\a$ in a $k+l$-dimensional space is
\eq{
\star_{k+l} \a = \frac{1}{l!}\epsilon_{m_1 ..m_l n_1...n_k} \a^{n_1 ... n_k} \;.
}
The Euclidean Hodge star for $Spin(d)$ gamma-matrices with $d$ even satisfies
\eq{
\g_{(k)} = (-1)^{\frac12 d (d-1)} (-1)^{\frac12 k (k-1)} \star_d \g_{(d-k)} \g_{(d)}
}
while the Lorentzian equivalent for $Spin(1,d-1)$ is given by
\eq{
\g_{(k)} = - (-1)^{\frac12 d (d-1)} (-1)^{\frac12 k (k-1)} \star_d \g_{(d-k)} \g_{(d)} \;.
}
The chirality matrix for either signature is defined as
\eq{
\g_{(d)} = \frac{1}{d!} \e^{M_1 ... M_d} \g_{M_1 ... M_d} \;.
}
Relevant (anti-)commutator identities for gamma-matrices which have been made use of are
\eq{
\begin{alignedat}{6}
\{\G^{NP}, \G_M\}        &= 2 \G_{M}^{\phantom{M}NP}
&& \quad &&   [\G^{MN}, \G_P]     &&= - 4 \delta^{[M}_P \G^{N]}\\
\{\G^{NP}, \G_{QM} \}    &= 4 \delta^{NP}_{QM} - 2 \G_{MQ}^{\phantom{MQ}NP}
&& \qquad &&  [\G^{MN}, \G_{PQ} ] &&= - 8 \delta^{[M}_{[P} \G^{N]}_{\phantom{B]}Q]} \\
\{\G^{NP}, \G_{QRM} \}   &= -12 \delta^{NP}_{QR}\G_M + 2 \G_{MQR}^{\phantom{MQR}NP}
&& \qquad &&  [\G^{MN}, \G_{PQRS} ] &&= - 16 \delta^{[M}_{[P} \G^{N]}_{\phantom{B]}QRS]}\\
 \{\G^{NP}, \G_{QRSM} \} &= 24 \delta^{NP}_{QR}\G_{MS} - 2 \G_{MQRS}^{\phantom{MQRS}NP}
&& \qquad && \G_{M_1 ... M_k} \G_N &&= \G_{M_1 ... M_k N} + k \G_{[M_1 ... M_{k-1}} \delta_{M_k] N} \;.
\end{alignedat}
}
Note that these are both dimension and signature independent. When splitting $Spin(1,9) \rightarrow Spin(1,5) \times Spin(4)$, we decompose the ten-dimensional gamma-matrices as
\eq{ \label{gammadecomp}
\G^m &= \obb \otimes \g_m \\
\G^a &= \check{\g}^a \otimes \g_{(4)}  \;.
}
Our spinor conventions are as follows. The charge conjugation matrix $C_{2n}$ in $2n$-dimensional spacetime is given (up to sign) by the anti-symmetrized product of $n$ spacelike gamma matrices. It satisfies
\eq{
C^{-1} = C^\dagger \;.
}
We work with hermitian spacelike gamma matrices and anti-hermitian timelike gamma matrices.
For Euclidean even-dimensional spaces, we define Majorana conjugation of Weyl spinors by
\eq{\label{ecc}
(\psi^\pm)^c =  C (\psi^\pm)^* \;.
}
For Lorentzian even-dimensional spaces, we define Majorana conjugation of spinors by
\eq{\label{lcc}
(\psi^\pm)^c = \G_0 C (\psi^\pm)^* \;.
}
For the specific dimensions we require, we find the following identities:

\subsection*{$Spin(4)$}
The charge conjugation matrix satisfies
\eq{
C_4^T = - C_4 \;, \qquad (C_4 \g_m)^T = - C_4 \g_m \;.
}
Spinors are pseudoreal, in the sense that Majorana conjugation does not change chirality, nor is it involutive:
\eq{
[(\t^\pm)^c]^c = - \t^\pm\;.
}
The chirality matrix satisfies
\eq{
\g_{(4)}^2  = 1 \;.
}

\subsection*{$Spin(1,5)$}
The charge conjugation matrix satisfies
\eq{
C_6^T =  C_6 \;, \qquad (C_6 \cg_a)^T = - C_6 \cg_a \;.
}
Spinors are pseudoreal, in the sense that Majorana conjugation does not change chirality, nor is it involutive:
\eq{
[(\t^\pm)^c]^c = - \t^\pm\;.
}
The chirality matrix satisfies
\eq{
\cg_{(6)}^2  = 1 \;.
}

\subsection*{$Spin(1,9)$}
The charge conjugation matrix satisfies
\eq{
C_{10}^T =  - C_{10} \;, \qquad (C_{10} \G_M)^T =  C_{10} \G_M \;.
}
Spinors are real, in the sense that Majorana conjugation does not change chirality and is involutive:
\eq{
(\ve^\pm)^c = \ve^\pm \;.
}

\section{Squaring linear supersymmetry}
\label{sec:Lich}

The integrability theorems are among of the standard tools used in analysing supersymmetric flux compactifications. We can use this tool to argue the merits of the proposal for six-dimensional linear supersymmetry made in subsection \ref{D5inst}. While the details vary widely depending on the details of the backgrounds, and the calculations can be involved, the basic idea is fairly simple: denoting the supersymmetry equations schematically as \eq{\label{schematic}
Q \ve= 0 \;,
}
one wants to use the ensuing condition $Q^2 \ve=0 $ to see which equations of motion follow automatically when \eqref{schematic} is satisfied, and end up with a set of (hopefully) simpler conditions that ensure the solution of the full theory given preservation of supersymmetry. These conditions universally involve Bianchi identities, and in some backgrounds a subset of equations of motion. Lichnerowicz formula is probably the simplest and best known of these theorems. For Levi-Civita connections the difference of squares of the Dirac operator and the covariant derivative is proportional to the Ricci scalar, i.e. the trace of the Einstein equation. As the vanishing of the action is often an equation of motion for supersymmetric theories, one can expect that, modulo vanishing on-shell terms, $Q^2 \sim \lcal$.\footnote{For extension of the Lichnerowicz formula to connection with torsion see \cite{bismut}; for recent applications (involving higher derivative terms) to heterotic strings and M-theory see \cite{cmtw} and \cite{cm} respectively.}  As we shall argue here, that differently from $Q \sin \cancel{\fcal}$, the choice of $Q$ given by the linear generators in \eqref{d=6nlinstanton} leads to a good six-dimensional Lagrangian for the gauge field (involving higher-derivative terms).

%and the bosonic equations of motion by
%\eq{\label{schematic2}
%\ecal = 0 \;.
%}
%then one can find that modulo (a lot of) details (such as Bianchi identities), $Q \ve=0 $ implies that the (bosonic) equations of motions are satisfied, hence one concludes that one way to find backgrounds is to just solve \eqref{schematic} instead. Very roughly speaking, this is due to the fact that $Q^2 \sim \ecal$. Inspired by this idea for closed strings, we will examine whether for open strings (more specifically, the supersymmetry of D-branes as defined by the Green-Schwartz operator) something similar occurs. As it turns out, we do not find that $Q^2 \sim \ecal$. Instead, $Q^2 \sim \lcal$, with $\lcal$ the Yang-Mills Lagrangian.

Let us start with a warm-up exercise and consider $d=4$ linear supersymmetry. Taking  $\ve_- = 0$ we expect to end up with HYM as described in section \ref{hym}.
The instanton equation, to which we refer in this context as the supersymmetry equation, is given by
\eq{
\cancel{\fcal} \ve_+^1 &= 0 \\
\left(\sqrt{ \det( \delta + g^{-1}\fcal)} - 1-\frac14 (\star \fcal)_{mn} \fcal^{mn} \g_{(4)}\right) \ve_+^1 &= 0 \;.
}
As described in section \ref{hym}, the two equations are equivalent. Let us consider $Q \equiv \cancel{\fcal}$.
Then one finds that
\eq{
Q^2 = \cancel{\fcal}^2 = \frac12 \fcal_{mn} \fcal_{pq}g^{mp}g^{nq}  + \frac14 \e^{mnpq} \fcal_{mn} \fcal_{pq} \g_{(4)} \;.
}
In order to get rid of $\g_{(4)}$, we would like to consider $|Q \ve|^2$. However, spinor calculus (or representation theory) leads to the conclusion that
$ \tilde{\ve^c} \ve= \ti{\ve} \ve = 0$. Instead, what we should do is consider the spinor decomposition $Spin(1,9) \rightarrow Spin(9)\rightarrow Spin(3) \otimes Spin(6)$,
\eq{
\ve_+^1 &= \left( \begin{array}{c} 1 \\ 0 \end{array} \right) \otimes \hat{\ve} = \left( \begin{array}{c} 1 \\ 0 \end{array} \right) \otimes \left( \theta^+ \otimes \eta^+ \right) \;.
}
Then we can consider $|Q \hat{\ve}|^2$, since $\hat{\ve}$ does have a non-trivial norm, which we fix as $g_{YM}^{-1}$. See \cite{kt} section 4.1 or \cite{lpt} section 3.1 for details.
We rescale the metric
\eq{
g  \rightarrow \sqrt{\frac{i \t}{8 \pi^2}} g
}
to find
\eq{
-\ti{\hat{\ve}}^c Q^2  \hat{\ve} &=  \lcal \equiv - \frac{1}{2g_{YM}} \fcal_{mn} \fcal^{mn} - \frac{i \theta }{32 g_{YM} \pi^2} \e^{mnpq} \fcal_{mn} \fcal_{pq} \;.
}
We thus  have  managed to construct the $d=4$ Yang-Mills Lagrangian by squaring the supersymmetry conditions of the D3-brane.

Encouraged by this result, let us try to extend this procedure to $d=6$. The linear supersymmetry equations are given by
\eq{\label{d=6lininst}
(\cancel{\fcal} - \rcal \g_{(6)}) \ve_+^1 &= 0 \\
( \sqrt{ \det( \delta + g^{-1}\fcal) } - \qcal \g_{(6)} - 1) \ve_+^1 &=  0 \;,
}
with $\qcal$ and $\rcal$ defined in \eqref{pqr}. In particular, it can be shown that
\eq{
 & \det ( \delta + g^{-1} \fcal)  = \\
 & 1 + \fcal^2 + \left(\frac{1}{2} (\fcal^2)^2 - \frac14 \Tr \fcal^4 \right) +
 \left( \frac{1}{3!} (\fcal^2)^3 - \frac14 \fcal^2 \Tr \fcal^4 - \frac16 \Tr \fcal^6 \right)  \;.
 }
For $d=4$ (the D3-brane), the two equations defining linear supersymmetry were equivalent. Here, this is not the case, and both operators are required in order to construct a combination yielding only scalar-like quantities.
%which might leave one to wonder which should be considered `the $Q$' to square. As it turns out, we will require both to obtain something sensible.
Let us consider $Q \equiv \cancel{\fcal} - \rcal \g_{(6)}f$. Then, we find that
\eq{
Q^2 = ( \cancel{\fcal} - \rcal )^2 = \frac12 \fcal_{mn} \fcal^{mn} + (1 + \frac12 \fcal_{mn} \fcal^{mn}) \frac14 \fcal_{mn}\fcal_{pq} \g^{mnpq} + \frac{1}{3!} (\frac12 \fcal_{mn} \fcal^{mn})^3
}
As the second term includes gamma-matrices, $Q^2$ is not a scalar. However, we now make use of the second equation of \eqref{d=6lininst} to rewrite this term. We then find
\eq{
- \tilde{\hat{\ve}}^c Q^2 \hat{\ve} &=  \lcal \equiv - \frac{1}{2 g_{YM}} \fcal_{mn} \fcal^{mn} - (1 + \frac12 \fcal_{mn} \fcal^{mn})
\frac{\sqrt{\det( \delta + g^{-1} \fcal)}-1}{g_{YM}} - \frac{1}{3!g_{YM}} (\frac12 \fcal_{mn} \fcal^{mn})^3 \;.
}
We would propose that this $d=6$ Lagrangian, which includes higher derivative terms, may be interesting for further study.

\section{Derivation of the Killing spinor equations}\label{ftcalc}
In this section, we give the derivation of the Killing spinor equations of four-dimensional $\ncal=8$ supergravity \eqref{n=4ks} (which leads to the simplified \eqref{n=4ks2}) and $\ncal=2$ supergravity  \eqref{n=1fs}, starting from the string theory supersymmetry condition \eqref{genfs} (or equivalently, \eqref{ft}). As can be seen, the starting point is almost purely the closed string supersymmetry condition, albeit in the basis  $\ve_\pm$ that is strongly inspired by kappa-symmetry. The only place where D3-brane data enters the computation is in the decomposition of the Killing spinors \eqref{ftksdecomp}: the internal spinors in the decompositions of both $\ve_+$ and $\ve_-$ need to be identical to allow for non-trivial worldvolume flux. This situation is in contrast with the situation without non-linear supersymmetries as described in \cite{triendl}, where $\G_0 \ve = \ve$ is a constraint on $\ve$ imposed by the D3-brane.

\subsection{$\ncal = 4$}
In order to obtain the $d=4$ Killing spinor equations with $\ncal=4$ linear supersymmetries, the starting point is the the pair of doublet equations equations \eqref{ft}. However, \eqref{ft} makes use of the unmodified external gravitino variation. Instead, we are interested in the modified external gravitino variation, \eqref{extgravshift}. There are two ways one can proceed. The first way is to compute the (anti-)commutators of the unmodified shifts, and then take a linear combination at the end. The second way is to first construct the differential operator associated to the susy variation of the modified external gravitino, and then compute the (anti-)commutators using this differential operator rather than $D_m$. This requires one to realize in advance that such a shift will be necessary for the representation theory to work out, and to know the correct coefficients beforehand, which would be possible by careful study of \cite{deWit:1986mz}. As we did not, we will instead demonstrate the computations using unmodified ten-dimensional fermions.

In all three cases, the computation is very similar, with the sole difference of using a different operator for the commutation relations. Let us start by demonstrating the unmodified external gravitino.  We consider
\eq{\label{ft12}
L^\pm_m \equiv \frac12 \{D_m, \G_0\} \ve_\pm  = \frac12 [D_m \G_0] \ve_\mp \equiv R^\mp_m \;,
}
where we have introduced the notation $L^\pm_m$, $R^\mp_m$ for the left- and right-hand side for convenience. We can work these out by making use of \eqref{eq:gravitinovar}. $L_m^+$ was worked out in \cite{triendl}, and leads to
\eq{\label{LHS}
\frac12\{D_m, \G_0\} \ve_+ &= \left(\nabla_m + \frac14 \o_{mab} \cg^{ab} - \frac18 i e^\phi \left( F_n \g_{(4)} + \frac{1}{48} \e^{abcdef} F_{cdefn}  \cg_{ab} \right)\g^n \g_m \right)\ve_+ \\
&+ \frac14\left( \left(H_{mna}\g_{(4)} - \frac12 i e^\phi ( F_{mna}  - \frac12 \e_{mnpq} F_a^{\phantom{a}pq} \g_{(4)}) \right)\g^n \cg^a
 + \frac{1}{12}ie^\phi F_{abc} \cg^{abc} \g_m  \right)  \pcal \ve_+ \;,
}
whereas the right-hand side $R_m^-$ is given by
\eq{\label{RHS}
\frac12 [D_m, \G_0] \ve_- =&
\frac12 \left( -  \o_{mna} \g^n \cg^a \g_{(4)}  + \frac14 i e^\phi  \left( \left( F_a  + F_{a1234} \g_{(4)} \right) \cg^a \g_m
 - \frac{1}{3!} F_{mnabc} \cg^{abc} \g^n \right)\right) \ve_- \\
&- \frac18 \left( H_{mnp}\g^{np} + H_{mab} \cg^{ab}  + \frac12 i e^\phi \left( F_{nab} \cg^{ab} \g_{(4)} - \frac13 \e_{npqr} F^{pqr}\right) \g^n \g_m\right) \pcal \ve_-\;.
}
Furthermore, one can obtain $L_m^-$ and $R_m^+$ by switching $\ve_\pm \longleftrightarrow \ve_\mp$ and $F \rightarrow - F$.
It will prove to be convenient to introduce $a_m, ..., d_m$ and $\ti{a}_m, .., \ti{d}_m$ as an intermediate step, defined by
\eq{\label{addef}
L^+_m &=  \left( a_m + b_m \g_{(4)} + c_m \pcal + d_m \g_{(4)} \pcal \right) \left(1 + \G_0\right) \ve_+^1 \\
R^-_m &= \left( \ti{a}_m + \ti{b}_m \g_{(4)} + \ti{c}_m \pcal + \ti{d}_m \g_{(4)} \pcal \right) \left(1 - \G_0\right) \ve_-^1 \;.
}
Comparing \eqref{addef} with \eqref{LHS} and \eqref{RHS}, we find the following expressions:
\eq{
a_m &= \left( \nabla_m + \frac14 \o_{mab} \cg^{ab} \right) + \left( - \frac{1}{4!2! 8} i e^\phi \e^{abcdef} F_{cdefn}\cg_{ab}\right)\g^n \g_m  \\
b_m &= - \frac{1}{8} i e^\phi F_n \g^n \g_m \\
c_m &= \left(- \frac{1}{8} i e^\phi F_{mna} \cg^a\right) \g^n  + \left(\frac{1}{48} i e^\phi F_{abc} \cg^{abc} \right) \g_m \\
d_m &= \left( \frac14 H_{mna} \cg^a - \frac{1}{16} i e^\phi \e_{mnpq} F^{pqa} \cg_a \right) \g^n
}
and
\eq{
\ti{a}_m &= \left( \frac18 i e^\phi F_a \cg^a \right) \g_m + \left(-\frac{1}{48} i e^\phi F_{mnabc} \cg^{abc}\right)\g^n \\
\ti{b}_m &= \left( - \frac12 \o_{mna} \cg^a \right) \g^n + \left( - \frac{1}{8} i e^\phi  F_{a1234} \cg^a \right) \g_m \\
\ti{c}_m &= \left( - \frac{1}{8} H_{mab} \cg^{ab} \right) + \left(\frac{1}{48} i e^\phi \e_{npqr} F^{pqr} \right) \g^n \g_m \\
\ti{d}_m &=  \left(- \frac{1}{16} i e^\phi  F_{abn} \cg^{ab}\right) \g^n \g_m - i \hcal_m + i \hcal_n \g^n \g_m \;,
}
where we have introduced
\eq{
\hcal_m = - \frac{1}{4!}i \e_{mnpq} H^{npq} \;.
}
We denote the doublet components as $L^+_m = (L^{+}_{1|m}, L^+_{2|m})$, such that \eqref{ft12} is equivalent to the equations
\eq{\label{ft3}
\frac12 (L_{1|m}^+ \pm i L_{2|m}^+) &= \frac12 (R_{1|m}^- \pm i R_{2|m}^-) \\
\frac12 (L_{1|m}^- \pm i L_{2|m}^-) &= \frac12 (R_{1|m}^+ \pm i R_{2|m}^+) \;.
}
Using the Killing spinor decomposition \eqref{ftksdecomp},
it easily follows that  (suppressing spinor-sum indices $\a$ and the explicit tensor product)
\eq{\label{Lterms}
\frac12 (L_{1|m}^+ + i L_{2|m}^+) &=  \left( a_m + b_m \right) \xi^+ \eta^+ + \left( c_m - d_m \right) \xi^- \eta^- \\
\frac12 (L_{1|m}^+ - i L_{2|m}^+) &=  \left( a_m - b_m \right) \xi^- \eta^- + \left( c_m + d_m \right) \xi^+ \eta^+
}
as well as
\eq{\label{Rterms}
\frac12 (R_{1|m}^- + i R_{2|m}^-)   &= \left( \ti{c}_m + \ti{d}_m \right) \xi^+ \z^+ + \left( \ti{a}_m - \ti{b}_m \right) \xi^- \z^- \\
\frac12 (R_{1|m}^- - i R_{2|m}^-)   &= \left( \ti{c}_m - \ti{d}_m \right) \xi^- \z^- + \left( \ti{a}_m + \ti{b}_m \right) \xi^+ \z^+ \;.
}
Inserting \eqref{Lterms} and \eqref{Rterms} into \eqref{ft3}, we find the following four equations; again, we note that the third and fourth equation can be obtained from the first and second by taking $\eta^\pm \leftrightarrow \z^\pm$ and $F \rightarrow - F$.
\begin{subequations}\label{n=4fsfull}
\al{
& \left(\nabla_m + A_m^{0\,+} +  (U_n^+ + i s_n) \g^n \g_m  \right) \eta^+ + \left( V_{mn}^+ \g^n +  t^+ \g_m +  i (X_{mn}^+ + Y_{mn}^+) \g^n  \right) \eta^-
\label{n=4fsfull1} \\
 &= - \left( \Delta_m^+ + i \hcal_m + (\Lambda_{n} + i \Xi_n^+ - i \hcal_n) \g^n \g_m  \right) \z^+
- \left( ( \Pi^+ + i \Theta^+) \g_m + (\Sigma_{mn}^+ + i \Upsilon_{mn}^+ ) \g^n  \right) \z^-  \nonumber\\
&\nonumber\\
& \left( \nabla_m + A_m^{0\, -} +  (U_n^- - i s_n) \g^n \g_m \right) \eta^- + \left( V_{mn}^- \g^n + t^- \g_m   - i (X_{mn}^- + Y_{mn}^-) \g^n \right) \eta^+
 \label{n=4fsfull2}\\
  &= - \left(  \Delta_m^- - i \hcal_m + (\Lambda_{n} - i \Xi_n^- + i \hcal_n)\g^n \g_m \right) \z^-
- \left( (\Pi^- - i \Theta^-) \g_m + (\Sigma_{mn}^-   - i \Upsilon_{mn}^-) \g^n \right) \z^+ \nonumber\\
&\nonumber \\
& \left(\nabla_m + A_m^{0\,+} -  (U_n^+ + i s_n) \g^n \g_m \right) \z^+ + \left( - V_{mn}^+ \g^n   - t^+ \g_m  + i (X_{mn}^+ - Y_{mn}^+) \g^n\right) \z^-
 \label{n=4fsfull3} \\
  &= - \left(  \Delta_m^+ + i \hcal_m + (-\Lambda_{n} - i \Xi_n^+ - i \hcal_n) \g^n \g_m  \right) \eta^+
- \left( -(\Pi^+ + i \Theta^+) \g_m + (- \Sigma_{mn}^+ + i \Upsilon_{mn}^+) \g^n  \right) \eta^- \nonumber\\
&\nonumber \\
& \left( \nabla_m + A_m^{0\, -} -  (U_n^-  - i s_n) \g^n \g_m \right) \z^- + \left( - V_{mn}^- \g^n - t^- \g_m - i (X_{mn}^- - Y_{mn}^-) \g^n  \right) \z^+
\label{n=4fsfull4}\\
&= - \left(  \Delta_m^- - i \hcal_m + (-\Lambda_{n} + i \Xi_n^-+ i \hcal_n) \g^n \g_m  \right) \eta^-
- \left( -(\Pi^- - i \Theta^-) \g_m  + (- \Sigma_{mn}^- - i \Upsilon_{mn}^- )\g^n  \right) \eta^+ \nonumber\;,
}
\end{subequations}
with fields
\eq{\label{n=4fields}
\begin{alignedat}{4}
A_m^{0 \, \pm} \xi^\pm    &=  \frac14 \o_{mab} \cg^{ab} \xi^\pm
&& \qquad \qquad & \Delta_m^\pm       \xi^\pm  &=    \frac18  H_{mab} \cg^{ab} \xi^\pm  \\
U_m^\pm \xi^\pm    &= -\frac{1}{384} i e^\phi \e_{abcdef} F_m^{\phantom{m}cdef}  \cg^{ab} \xi^\pm
&&\qquad  \qquad &  \Xi_m^\pm         \xi^\pm  &=    \frac{1}{16}   e^\phi   F_{mab} \cg^{ab} \xi^\pm   \\
s_m               &= -\frac18  e^\phi  F_m
&& \qquad \qquad & \Lambda_{m}                &=  - \frac{1}{48} i e^\phi  \e_{mnpq} F^{npq}  \\
V_{mn}^\pm \xi^\pm &= -\frac18 i e^\phi  F_{mna}\cg^a \xi^\mp
&& \qquad \qquad & \hcal_m                    &=  - \frac{1}{4!} i         \e_{mnpq} H^{npq}  \\
X_{mn}^\pm \xi^\pm &=  \frac14 i H_{mna}  \cg^a \xi^\mp
&& \qquad \qquad & \Pi^\pm            \xi^\pm  &=  - \frac18 i e^\phi  F_a       \cg^a  \xi^\mp\\
Y_{mn}^\pm \xi^\pm &=  \frac{1}{16}  e^\phi \e_{mnpq} F_a^{\phantom{a}pq} \cg^a \xi^\mp
&& \qquad \qquad & \Theta^\pm         \xi^\pm  &=  - \frac18   e^\phi  F_{a1234} \cg^a  \xi^\mp \\
t^\pm \xi^\pm      &=  \frac{1}{48} i e^\phi F_{abc} \cg^{abc}\xi^\mp
&& \qquad \qquad & \Upsilon_{mn}^\pm  \xi^\pm  &=    \frac12 i \o_{mna} \cg^a \xi^\mp \\
\phantom{a} &\phantom{=} && \qquad \qquad & \Sigma_{mn}^\pm    \xi^\pm  &=    \frac{1}{48} i e^\phi F_{mnabc} \cg^{abc}\xi^\mp \;.
\end{alignedat}
}
Note that most fields are determined by eigenvalue equations for the internal spinors $\xi^\pm$; depending on the internal manifold, it may well be that some of these only admit trivial solutions.

In the case $\ve_- = 0$ (which is implied by taking $\fcal=0$), the RHS of the \eqref{n=4fsfull1}, \eqref{n=4fsfull2} are trivial, whereas it is the LHS of \eqref{n=4fsfull3} and \eqref{n=4fsfull4} that vanish. Thus, all fields decouple. The Killing spinor equation of $\ncal=4$ conformal supergravity is exactly given by the first two lines in this case.\footnote{See \cite{triendl}, eq. (3.6), (3.7) for comparison.}

Next, let us recombine the four equations \eqref{n=4fsfull} into two doublet matrix equations. We take  \eqref{n=4fsfull1} together with \eqref{n=4fsfull3} which leads to the first doublet equation, while \eqref{n=4fsfull2} together with \eqref{n=4fsfull4} lead to another. These doublet equations are given by
\eq{\label{n=4extgrav}
(\ti{\nabla}_m + A_m^+) \l^+ + \ti{T}_{mn}^+ \g^n \l^- + \g_m (\ti{K}^+ \l^- - \ti{A}_n^+ \g^n \l^+) &= 0 \\
(\ti{\nabla}_m + A_m^-) \l^- + \ti{T}_{mn}^- \g^n \l^+ + \g_m (\ti{K}^- \l^+ - \ti{A}_n^- \g^n \l^-) &= 0 \;,
}
with fields given by
\eq{
\ti{A}_m^\pm &= \left( \begin{array}{cc}
A_m^{0\pm}  +  2(U_m^\pm \pm i s_m)  &
\Delta_m^\pm \mp i \hcal_m  +   2(\L_m \pm i \Xi^\pm_m) \\
\Delta_m^\pm \mp i \hcal_m  -   2(\L_m \pm i \Xi^\pm_m) &
A_m^{0\pm} -   2(U_m^\pm \pm i s_m)
\end{array} \right)
}

and
\eq{
\ti{T}_{mn}^\pm &= \left( \begin{array}{cc}
\pm i X_{mn}^\pm + V_{mn}^\pm \pm i  Y_{mn}^\pm &
\pm i \Upsilon_{mn}^\pm + \Sigma_{mn}^\pm \\
\pm i \Upsilon_{mn}^\pm - \Sigma_{mn}^\pm &
\pm i X_{mn}^\pm - V_{mn}^\pm \mp i  Y_{mn}^\pm
\end{array} \right)\;.
}
The terms appearing in the conformal transformation are given by
\eq{
\ti{K}^\pm &= \left(\begin{array}{cc}
t^\pm &
\Pi^\pm \pm i \Theta^\pm \\
- \Pi^\pm \mp i \Theta^\pm &
- t^\pm
\end{array}\right)
}
and
\eq{
\ti{A}_m^\pm &= \left( \begin{array}{cc}
     (U_m^\pm \pm i s_m)  &
\mp i \hcal_m  +   (\L_m \pm i \Xi^\pm_m) \\
\mp i \hcal_m  -   (\L_m \pm i \Xi^\pm_m) &
-   (U_m^\pm \pm i s_m)
\end{array} \right) \;.
}
This is the final result for the unmodified external gravitino equation.

The way to construct the dilatino equation and the trace of the internal gravitino is entirely analogous to the procedure above:
\begin{itemize}
\item Construct the analogues $L^\pm$, $R^\mp$ as defined in \eqref{ft12}, replacing $D_m$ either by $D$ or $\G^a D_a$, as defined in \eqref{eq:gravitinovar}, \eqref{eq:dilatinovar}.
\item Read off definitions of $a, ..., d$ and $\ti{a}, ... \ti{d}$ which are defined as in \eqref{addef} for $L^\pm$, $R^\mp$.
\item Compute the analogue of \eqref{ft3} to obtain the four equations that are the analogue  of \eqref{n=4fsfull}.
\item Recombine these four into two doublet equations. Read off the fields.
\end{itemize}
The results of each of these intermediate steps are as follows. For the dilatino, we find that
\eq{\label{d=10dilatino1}
L^+ &=  \left( \p_a \phi \g_{(4)} + i e^\phi F_a \right) \cg^a \ve_+ \\
& \phantom{=+} + \frac{1}{12} \left( 3 H_{mab} \cg^{ab} \g^m + \e_{mnpq} H^{npq} \g^m \g_{(4)}
              + i e^\phi \left(3 F_{mab} \cg^{ab} \g^m \g_{(4)} + \e_{mnpq} F^{npq} \g^m  \right)  \right) \pcal \ve_+  \\
R^- &= \left( - \p_m \phi + i e^\phi F_m \g_{(4)} \right) \g^m \ve_- \\
& \phantom{=+} + \frac{1}{12} \left( \left(-H_{abc} \g_{(4)} + i e^\phi F_{abc} \right) \cg^{abc}
+ 3 \left(- H_{mna} \g_{(4)} + i e^\phi F_{mna} \right) \g^{mn} \cg^{a} \right) \pcal \ve_- \;.
}
From this, we see that
\eq{\label{dilatonabcd}
\begin{alignedat}{6}
a &= i e^\phi F_a \cg^a  & \qquad \qquad & \ti{a} &=& - \p_m \phi \g^m \\
b &= \p_a \phi \cg^a     & \qquad \qquad & \ti{b} &=&  i e^\phi F_m \g^m \\
c &= \frac{1}{12} \left( i e^\phi \e_{mnpq} F^{npq} + 3          H_{mab} \cg^{ab} \right) \g^m  & \qquad \qquad &
\ti{c} &=&  \frac{1}{12} i e^\phi \left( F_{abc} \cg^{abc} + 3 F_{mna} \cg^{a} \g^{mn} \right) \\
d &= \frac{1}{12} \left(          \e_{mnpq} H^{npq} + 3 i e^\phi F_{mab} \cg^{ab} \right) \g^m  & \qquad \qquad &
\ti{d} &=& - \frac{1}{12}          \left( H_{abc} \cg^{abc} + 3 H_{mna} \cg^{a} \g^{mn} \right) \;.
\end{alignedat}
}
Inserting these in the analogue of \eqref{ft12} yields
\eq{
& \left(-8 \Pi^- + 2 i \dcal^- \right) \eta^+ + 2 \left(\Delta_m^- - 2 \L_m - 2 i \Xi_m^- - i \hcal_m \right) \g^m \eta^- \\
&=  \left( 4 t^- + 2 i G^- + (- 2 V_{mn}^-  +  i X^-_{mn} ) \g^{mn} \right) \xi^+ + \left(- \p_m \phi +  8 i s_m \right)\g^m\xi^- \\
&\\
& \left(-8 \Pi^+ - 2 i \dcal^+ \right) \eta^- + 2 \left(\Delta_m^+ - 2 \L_m + 2 i \Xi_m^+ + i \hcal_m \right) \g^m \eta^+ \\
&=  \left( 4 t^+ - 2 i G^+ + (- 2 V_{mn}^+ -  i X^+_{mn} ) \g^{mn} \right) \xi^- + \left(- \p_m \phi - 8 i s_m \right)\g^m\xi^+ \\
&\\
& \left(8 \Pi^- + 2 i \dcal^- \right) \xi^+ + 2 \left(\Delta_m^- + 2 \L_m + 2 i \Xi_m^- - i \hcal_m \right) \g^m \xi^- \\
&=  \left(-  4 t^- + 2 i G^- + (2 V_{mn}^- +  i X^-_{mn} ) \g^{mn} \right) \eta^+ + \left(- \p_m \phi - 8 i s_m \right)\g^m\eta^- \\
&\\
& \left(8 \Pi^+ - 2 i \dcal^+ \right) \xi^- + 2 \left(\Delta_m^+ + 2 \L_m - 2 i \Xi_m^+ + i \hcal_m \right) \g^m \xi^+ \\
&=  \left(-  4 t^+ - 2 i G^+ + (2 V_{mn}^+ -  i X^+_{mn} ) \g^{mn} \right) \eta^- + \left(- \p_m \phi + 8 i s_m \right)\g^m\eta^+ \;.
}
where we have introduced the following fields that did not yet appear in the external gravitino:
\eq{\label{n=4dilatinofields}
G^\pm \xi^\pm &=  - \frac{1}{4!} i H_{abc} \cg^{abc} \xi^\mp \\
\dcal^\pm \xi^\pm & = \frac12 i \p_a \phi \cg^a \xi^\mp \;.
}
Combining the first and third, and the second and fourth, of the above equations then leads to equations
 \eq{\label{n=4dilatino}
\kcal^+ \l^- - \tcal^{+\m}_{mn}  \g^{mn} \l^- + \acal^+_m   \g^m \l^+ &= 0 \\
\kcal^- \l^+ - \tcal^{-\m}_{mn}  \g^{mn} \l^+ + \acal^-_m   \g^m \l^- &= 0 \;.
}
The explicit expressions for the fields can be read off to find
\eq{
\kcal^\pm &= 2 \left(\begin{array}{cc}
\pm i G^\pm + 2 t^\pm &
\mp i \dcal^\pm + 4 \Pi^\pm \\
\mp i \dcal^\pm - 4 \Pi^\pm &
\pm i G^\pm - 2 t^\pm
\end{array} \right)
}
and
\eq{
\acal_{m}^\pm &= \p_m \phi+ 2  \left(\begin{array}{cc}
\mp 4i s_m &
\Delta^\pm_m \pm i \hcal_m + 2 ( \Lambda_m \mp i \Xi^\pm_m ) \\
\Delta^\pm_m \pm i \hcal_m - 2 ( \Lambda_m \mp i \Xi^\pm_m ) &
\pm 4i s_m
\end{array} \right)
}
and
\eq{
\tcal_{mn}^\pm &= \left(\begin{array}{cc}
   \pm i X_{mn}^\pm - 2 V_{mn}^\pm & 0 \\
0& \pm i X_{mn}^\pm + 2 V_{mn}^\pm
\end{array} \right)\;.
}
Lastly, let us now give the expressions for the trace of the internal gravitino equations. The ten-dimensional internal gravitino equations are given by
\eq{
\frac12 \{ D_a, \G_0 \} \ve_+ =& \left(
\nabla_a^{(6)} + \frac14 \o_{amn} \g^{mn}
- \frac18 i e^\phi   \left( \left(F_b \g_{(4)}  + F_{b1234} \right) \cg^b  + \frac{1}{4!} F_{mnbcd}\cg^{bcd}  \g^{mn} \g_{(4)} \right) \cg_a \right) \ve_+ \\
&+ \frac14 \left(- H_{abm} \cg^b \g^m \g_{(4)}
- \frac{1}{4} i e^\phi \left(F_{mbc} \cg^{bc} - \frac13 \e_{mnpq} F^{npq} \g_{(4)} \right)  \g^m \cg_a \right) \pcal \ve_+
}
and
\eq{
\frac12 [D_a, \G_0] \ve_- =& \left(
- \frac12 \o_{abm} \cg^b \g^m \g_{(4)}
+ \frac18 i e^\phi \left( F_m + \frac{1}{4!2} \e_{bcdefg} F_m^{\phantom{m}bcde} \cg^{fg} \g_{(4)} \right) \g^m \cg_a \right) \ve_- \\
&+ \frac18 \left( H_{abc} \cg^{bc} + H_{amn} \g^{mn}
+ \frac{1}{3!} i e^\phi   \left( F_{bcd} \cg^{bcd} + 3 F_{mnb} \g^{mn} \cg^b \right)\g_{(4)}  \cg_a \right) \pcal \ve_-\;.
}
Taking the gamma-trace, we obtain the analogue of \eqref{dilatonabcd}, denoted with a subscript $\Psi$ to make the distinction clear:
\eq{
\begin{alignedat}{6}
a_\Psi &= \frac12 i e^\phi F_a \cg^a
     & \qquad \qquad & \ti{a}_\Psi &=& - \frac12 \o_{abm} \cg^a \cg^b \g^m - \frac{1}{4!8} i e^\phi \e_{abcdef} F_m^{\phantom{m}cdef} \cg^{ab} \g^m \\
b_\Psi &= \cancel{\nabla}^6 + \frac14 \o_{amn} \cg^a \g^{mn} + \frac12 i e^\phi F_{a1234} \cg^a
     & \qquad \qquad & \ti{b}_\Psi &=& \frac34 i e^\phi F_m \g^m \\
c_\Psi &= \frac{1}{8} \left( i e^\phi \e_{mnpq} F^{npq} + 2          H_{mab} \cg^{ab} \right) \g^m
     & \qquad \qquad & \ti{c}_\Psi &=&  \frac{1}{4} i e^\phi F_{mna} \cg^{a} \g^{mn}  \\
d_\Psi &= \frac{1}{8}  i e^\phi F_{mab} \cg^{ab} \g^m
     & \qquad \qquad &  \ti{d}_\Psi &=& - \frac18   \left( H_{abc} \cg^{abc} +  H_{mna} \cg^{a} \g^{mn} \right) \;.
\end{alignedat}
}
The resulting doublet equations can be expressed as:
\eq{\label{n=4intgrav}
\ch{K}^+ \l^- + \ch{T}_{mn}^+ \g^{mn} \l^- + \ch{A}_m^+ \g^m \l^+ -  \cancel{\nabla}^{6\pm} \s_1 \l^- &= 0\\
\ch{K}^- \l^+ + \ch{T}_{mn}^- \g^{mn} \l^+ + \ch{A}_m^- \g^m \l^- +  \cancel{\nabla}^{6\pm} \s_1 \l^+ &= 0\;,
}
with the fields defined as follows:
\eq{
\ch{K}^\pm \xi^\pm &=  \left(\begin{array}{cc}
\pm 3 i G &
4 \left( \Pi^\pm \mp i \Theta^\pm \right)\\
- 4 \left( \Pi^\pm \mp i \Theta^\pm \right) &
\pm 3 i G
\end{array} \right) \xi^\mp
}
and
\eq{
\ch{A}^\pm_m \z^\pm &=  \frac12 \o_{abm} \cg^{a}\cg^b \xi^\pm +
2 \left(\begin{array}{cc}
U_m^\pm   \mp  3 i s_m &
\Delta_m^\pm + 3 \L_m \mp i \Xi^\pm \\
\Delta_m^\pm - 3 \L_m \pm i \Xi^\pm &
- U_m^\pm \pm 3 i s_m
\end{array}\right) \xi^\pm
}
and
\eq{
\ch{T}_{mn}^\pm \xi^\pm =&
\mp \frac14 \o_{amn} \cg^a \s_1 \z^\pm +
\left(\begin{array}{cc}
\pm \frac12 i X_{mn}^\pm - 2 V_{mn}^\pm &
0\\0 &
\pm \frac12 i X_{mn}^\pm + 2 V_{mn}^\pm
\end{array}\right) \xi^\mp \;.
}
The term "$\cancel{\nabla}^{6\pm}$" is symbolic for the terms discussed in \eqref{torsion}.
We will not require the traceless part of the internal gravitino equation.

Having computed these (anti-)commutators of the external gravitino equation \eqref{n=4extgrav}, the dilatino equation \eqref{n=4dilatino} and the trace of the internal gravitino \eqref{n=4intgrav}, we add the three of them with relative weights $(1, -\frac12 \g_m, \frac12 \g_m)$, which is equivalent to making use of the variation of the modified external gravitino\eqref{extgravshift}. The result is given in \eqref{n=4ks} with fields given in  \eqref{n=4fields1}, \eqref{n=4fields2}, \eqref{n=4fields3}.

\subsection{$\ncal=1$}
In order to derive the Killing spinor equations \eqref{n=1fs} for $\ncal=1$ field theory coupled to $\ncal =2$ supergravity, we need to break to $1/4$ of the supersymmetry. In order to do so, we will use \eqref{g2chi2} to construct projection operators with which we will act on both LHS and RHS of \eqref{ft3}. In particular, we define the projection operators
\eq{\label{proj}
\pi^\pm &\equiv  \frac14 (1 \pm i \cg_{89}) (1  \pm i \cg_{67})\;.
}
Using these projection operators, from \eqref{Lterms} we deduce that
\eq{\label{n=1lhs}
\pi^- \left(\frac12 (L_{1|m}^+ + i L_{2|m}^+)\right)   &=
- \frac14 \left( \{ \{a_m +  b_m, \cg_{67} \}, \cg_{89} \} \xi^+ \eta^+ + [[c_m  - d_m, \cg_{67}], \cg_{89} ] \xi^- \eta^- \right) \\
\pi^+ \left(\frac12 (L_{1m}^+ - i L_{2|m}^+ )\right)  &=
- \frac14 \left( \{ \{a_m -  b_m, \cg_{67} \}, \cg_{89} \} \xi^- \eta^- + [[c_m  + d_m, \cg_{67}], \cg_{89} ] \xi^+ \eta^+ \right) \;.
}
and
\eq{\label{n=1rhs}
\pi^- \left(\frac12 (R_{1|m}^- + i R_{2|m}^-) \right)   &=
- \frac14 \left( \{ \{\ti{c}_m + \ti{d}_m, \cg_{67} \}, \cg_{89} \} \xi^+ \z^+ + [[\ti{a}_m  - \ti{b}_m, \cg_{67}], \cg_{89} ] \xi^- \z^- \right) \\
\pi^+ \left(\frac12 (R_{1|m}^- - i R_{2|m}^- ) \right) &=
- \frac14 \left( \{ \{\ti{c}_m - \ti{d}_m, \cg_{67} \}, \cg_{89} \} \xi^- \z^- + [[\ti{a}_m  + \ti{b}_m, \cg_{67}], \cg_{89} ] \xi^+ \z^+ \right) \;.
}
The (anti-)commutators can be worked out by making use of
\eq{\label{Xcom2}
-\frac14 \{\{X_{ab} \cg^{ab}, \cg_{67} \}, \cg_{89} \}\xi^\pm  &= \pm 2!i  \left(i X_{05}  + X_{67}  + X_{89} \right) \xi^\pm \\
-\frac14 [[X_{a} \cg^{a}, \cg_{67}], \cg_{89}]\xi^\pm &=  0\\
-\frac14 [[X_{abc} \cg^{abc}, \cg_{67}], \cg_{89}]\xi^\pm &=
                                        - 3!\Big(  X_{068} - X_{079} + i X_{569} + i X_{578} \\
&\phantom{=} \mp \left(X_{568} - X_{579} + i X_{069} + i X_{078}\right)             \Big) \cg_{068}\xi^\pm \;,
}
which are obtained making use of \eqref{g2chi2}. We thus see precisely how the $\ncal = 8$ fields should be projected down to $\ncal=2$ fields with Poincar\'{e} symmetry broken by the $D7$-branes. Using this, one can go through the same motions as before, construct sets of four equations just as in \eqref{n=4fsfull} for unmodified external gravitino, dilatino and internal gravitino trace and then take the proper linear combination. Alternatively,  or one can just straight away project the result given in \eqref{n=4ks}. The latter is the more efficient way.

Nevertheless, we do wish to show the equations for the unmodified external gravitino. They are given by
\eq{\label{fsn=1}
& \left( \nabla_m + i A_m^0 + i V_n \g^n \g_m  \right) \eta^+ - M^+ \g_m \eta^- \\
&= - \left(  i A_m^1  + ( \frac12 A_n^2 - i \hcal_n) \g^n \g_m   \right) \z^+ - T_{mn}^+ \g^n \z^-\\
&\\
& \left( \nabla_m - i A_m^0 - i V_n \g^n \g_m  \right) \eta^- - M^- \g_m \eta^+ \\
&= - \left(- i A_m^1  +  (\frac12 A_n^2 + i \hcal_n) \g^n \g_m   \right) \z^- - T_{mn}^- \g^n \z^+ \\
&\\
& \left( \nabla_m + i A_m^0 - i V_n \g^n \g_m  \right) \z^+ + M^+ \g_m \z^- \\
&= - \left( i A_m^1  +  (- \frac12 A_n^2 - i \hcal_n) \g^n \g_m   \right) \eta^+ + T_{mn}^+ \g^n \eta^-\\
&\\
& \left( \nabla_m - i A_m^0 + i V_n \g^n \g_m  \right) \z^- + M^- \g_m \z^+ \\
&= -\left( -i A_m^1  +  (-\frac12 A_n^2 + i \hcal_n) \g^n \g_m   \right) \eta^- + T_{mn}^- \g^n \eta^+ \;.
}
with the fields defined as in \eqref{n=2fields1}, \eqref{n=2fields2}, \eqref{n=2fields3}, and where we have relabeled
\eq{\label{n=1fields}
V_m        = \frac12 A_m^3 \;, \qquad M^\pm      =  - M^{\pm1}\;.
}
The reason why we feel the need to pester the reader with more lengthy equations is to draw attention to the left-hand sides of the first two equations: setting $\ve_- = 0$ (i.e., $\z^\pm = 0$), one recovers precisely the gravitino variations of 16/16 supergravity as was found in \cite{triendl}. Thus, one now sees how this is embedded in our final result \eqref{n=1fs}.

\section{Breaking supersymmetry using D7-branes with flux}\label{d7}
 In this appendix, we will demonstrate why using D7-branes with non-trivial worldvolume flux to break the $\ncal=4$ Killing spinor equations down to $\ncal=1$ is problematic in the presence of non-trivial worldvolume flux.

Using similar notation as in section \ref{sec:n=1}, we denote the kappa-symmetry operators of the D3-brane with worldvolume flux and the D7-branes without worldvolume flux as respectively $\G^3$, $\G^7_0$, $\G^{\ti{7}}_0$. Then supersymmetry of all three branes is equivalent to enforcing
\eq{
\G^7_0 \ve= \G^{\ti{7}}_0 \ve = \G^3 \ve = \ve \;.
}
Taking into account our ansatze for the worldvolume flux, the kappa-symmetry operator for the Lorentzian D7-branes is given by
\eq{
\G^7_0 &= - i \g_{0567} \G^3_0 \\
\G^{\ti{7}}_0 &= - i \g_{0589} \G^3_0 \;.
}
In order to make computations doable, we will assume that the supersymmetry of the D7-branes is `aligned' with the projection along $\G^3_0$, that is to say, we assume
\eq{\label{ansatz}
\G^7_0 \ve_\pm = \ve_\pm \;.
}
Let us decompose $\ve_\pm^1$ as follows:
\eq{
\ve_+^1  &= \xi^+_\a \otimes \eta^+_\a + \xi^-_\a \otimes \eta^-_\a  \\
\ve_-^1  &= \chi^+_\a \otimes \z^+_\a + \chi^-_\a \otimes \z^-_\a \;.
}
Then making use of the fact that  $\G_{(10)} \ve\equiv \g_{(4)} \cg_{(6)} \ve= \ve$, and our assumption \eqref{ansatz}, it follows that
\eq{\label{g2chi1}
\begin{alignedat}{6}
\cg_{05} \xi^\pm &= \pm  \xi^\pm&  \qquad \qquad  &\cg_{05} \chi^\pm &=& \pm  \chi^\pm  \\
\cg_{67} \xi^\pm &= \pm i\xi^\pm&  \qquad \qquad  &\cg_{67} \chi^\pm &=& \mp i\chi^\pm\\
\cg_{89} \xi^\pm &= \pm i\xi^\pm&  \qquad \qquad  &\cg_{89} \chi^\pm &=& \mp i\chi^\pm\;.
\end{alignedat}
}
From this we deduce that, up to constant complex coefficients, the spinors must satisfy
\eq{\label{zetasol}
\xi^\pm &= \cg_{68} \chi^\pm \\
\xi^-   &= \cg_{068} \xi^+
}
We can absorb the constants into respectively, $\z^\pm$, $\eta^-$ and so we will set them to one without loss of generality. We now conclude that   $\chi^\pm \neq \xi^\pm$, contradicting our ansatz for the $\ncal =4$ case. However, this is problematic. Our analysis of the instanton equations in section \ref{nli} demonstrated exactly that supersymmetry of the D3-brane required the internal components of $\ve_\pm$ to be equivalent in order to admit a non-trivial worldvolume flux. For $\xi^\pm \neq \chi^\pm$, we find that $\fcal = 0 \implies \ve_- = 0$. This thus leads to the case already analyzed in \cite{triendl}.

\end{appendices}

\end{document}